\documentclass[english,nodate]{elsarticle}

\usepackage[utf8]{inputenc}
\usepackage[active]{srcltx}
\usepackage{array}
\usepackage{float}
\usepackage{multirow}
\usepackage{amsmath}
\usepackage{amssymb}
\PassOptionsToPackage{normalem}{ulem}
\usepackage{ulem}
\usepackage{natbib}
\makeatletter

\newcommand{\noun}[1]{\textsc{#1}}

\floatstyle{ruled}
\newfloat{algorithm}{tbp}{loa}
\providecommand{\algorithmname}{Algorithm}
\floatname{algorithm}{\protect\algorithmname}

\usepackage[usenames,dvipsnames]{pstricks}
 \usepackage{epsfig}
 \usepackage{pst-grad} 
 \usepackage{pst-plot} 

\usepackage{amsfonts}
\usepackage{algorithm}
\usepackage{algorithmic}

\usepackage{babel}

\newtheorem{theorem}{Theorem}

\newtheorem{coro}[theorem]{Corollary}

\newtheorem{example}{Example}

\newcommand{\proof}{{\bf Proof.}\ }
\newcommand{\kommentar}[1]{}

\def\real{\hbox{\rm\vrule\kern-1pt R}}
\def\nat{\hbox{\rm\vrule\kern-1pt N}}
\def\l{\ell}

\def\eps{\varepsilon}

\def\d{d}
\def\f{f}
\def\edv{ed^{1}}
\def\ed2v{ed^{2}}
\def\ed{ed}

\global\long\def\spg{{\sc Shortest Path Game}}
\global\long\def\spgd{{\sc Shortest Path Game}}

\global\long\def\geo{{\sc Geography}}
\global\long\def\vgeo{{\sc Vertex Geography}}
\global\long\def\psp{{\sf PSPACE}-complete}

\def\opath{{\em spe}-path}

\makeatother

\begin{document}


\title{On the Shortest Path Game \tnoteref{titlenote}}
\tnotetext[titlenote]{
Andreas Darmann was supported by the Austrian Science Fund (FWF):
{[}P 23724-G11{]}.
Ulrich Pferschy and Joachim Schauer were supported by the Austrian
Science Fund (FWF): {[}P 23829-N13{]}.
}

\author{Andreas Darmann}
\address{Institute of Public Economics, University of Graz,\\
Universitaetsstrasse~15, 8010 Graz, Austria,\\
\texttt{andreas.darmann@uni-graz.at}}

\author{Ulrich Pferschy}

\author{Joachim Schauer}
\address{Department of Statistics and Operations Research, University of Graz,\\
Universitaetsstrasse~15, 8010 Graz, Austria,\\
\texttt{\{pferschy, joachim.schauer\}@uni-graz.at}}

\date{}


\begin{abstract}
In this work we address a game theoretic variant of the shortest path
problem, in which two decision makers (players)
move together along the edges of a graph from a given starting vertex
to a given destination.
The two players take turns in deciding in each vertex which edge to traverse next.
The decider in each vertex also has to pay the cost of the chosen edge.
We want to determine the path where each player minimizes its costs
taking into account that also the other player acts in a selfish and rational way.
Such a solution is a subgame perfect equilibrium and
can be determined by backward induction in the game tree of
the associated finite game in extensive form.

We show that the decision problem associated with such a path is PSPACE-complete even
for bipartite graphs both for the directed and the undirected version.
The latter result is a surprising deviation from the complexity
status of the closely related game \geo.

On the other hand, we can give polynomial time algorithms for
directed acyclic graphs and for cactus graphs even in the undirected case.
The latter is based on a decomposition of the graph into components
and their resolution by a number of fairly involved dynamic programming arrays.
Finally, we give some arguments about closing the gap of the complexity status
for graphs of bounded treewidth.
\end{abstract}


\maketitle


\section{Introduction}
\label{sec:intro}

We consider the following game on a graph:
There is a directed graph $G=(V,A)$ given with vertex set $V$ and arc set $A$
with positive costs $c(u,v)$ for each arc $(u,v)\in A$
and two designated vertices $s,t\in V$.
The aim of \spg\ is to find a directed path from $s$ to $t$
in the following setting:
The game is played by two players $A$ and $B$
who have full knowledge of the graph.
They start in $s$ and always move together along arcs of the graph.
In each vertex the players take turns to select the
next vertex to be visited among all neighboring vertices of the current
vertex with player $A$ taking the first decision in $s$.
The player deciding in the current vertex also has to pay the cost
of the chosen arc.
Each player wants to minimize the total arc costs it has to pay.
The game continues until the players reach the destination vertex $t$.\footnote{
We will always assume that a path from $s$ to $t$ exists.}
Later, we will also consider the same problem on an undirected
graph $G=(V,E)$ with edge set $E$ which is quite different in several aspects.

We will impose two restrictions on the setting in order to ensure that vertex $t$ 
is in fact reachable and to guarantee finiteness of the game.
Even in a connected graph, it is easy to see that the players
may get stuck at some point and reach a vertex where the current player
has no emanating arc to select. To avoid such a deadlock, we restrict
the players in every decision to choose an arc (or edge, in the undirected case)
which still permits a feasible path from the current vertex
to the destination $t$ (which is computationally easy to check).
\begin{quote}
\textbf{(R1)} No player can select an arc which does not permit
a path to vertex $t$.
\end{quote}
In the undirected case, it could be argued that for a connected graph
this restriction is not needed since the players can always
leave dead ends again by going back the path they came.
Clearly, this would imply cycles of even length. 
Such cycles, however, may cause the game to be infinite as illustrated by
the following example.

\begin{example}\label{ex:infinite}
Consider the graph depicted below.
Player $B$ has to decide in vertex $v$ whether to pay the cost $M \gg 2$
or enter the cycle of length $3$.
In the latter case, the players move along the cycle and then $A$ has to decide
in $v$ with the same two options as before for player $B$.
In order to avoid paying $M$ both players may choose to enter the cycle
whenever it is their turn to decide in $v$ leading to an infinite game.
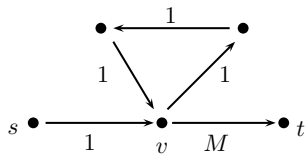
\begin{center}
\scalebox{0.9} 
{
\begin{pspicture}(0,-1.0829687)(5.0628123,1.0829687)
\psline[linewidth=0.03cm,arrowsize=0.05291667cm 2.0,arrowlength=1.4,arrowinset=0.4]{->}(0.7809375,-0.61546874)(2.3809376,-0.61546874)
\psline[linewidth=0.03cm,arrowsize=0.05291667cm 2.0,arrowlength=1.4,arrowinset=0.4]{->}(2.65,-0.61546874)(4.1809373,-0.61546874)
\psline[linewidth=0.03cm,arrowsize=0.05291667cm 2.0,arrowlength=1.4,arrowinset=0.4]{->}(2.5809374,-0.41546875)(3.5809374,0.58453125)
\psline[linewidth=0.03cm,arrowsize=0.05291667cm 2.0,arrowlength=1.4,arrowinset=0.4]{->}(3.5,0.78453124)(1.7809376,0.78453124)
\psline[linewidth=0.03cm,arrowsize=0.05291667cm 2.0,arrowlength=1.4,arrowinset=0.4]{->}(1.7809376,0.58453125)(2.3809376,-0.41546875)
\psdots[dotsize=0.16](0.6,-0.61546874)
\psdots[dotsize=0.16](2.5,-0.61546874)
\psdots[dotsize=0.16](3.7,0.78453124)
\psdots[dotsize=0.16](1.6,0.78453124)
\psdots[dotsize=0.16](4.3,-0.61546874)
\usefont{T1}{ptm}{m}{n}
\rput(1.4323437,-0.90546876){$1$}
\usefont{T1}{ptm}{m}{n}
\rput(1.6323438,0.09453125){$1$}
\usefont{T1}{ptm}{m}{n}
\rput(3.4323437,0.09453125){$1$}
\usefont{T1}{ptm}{m}{n}
\rput(2.6323438,0.97){$1$}
\usefont{T1}{ptm}{m}{n}
\rput(3.3023438,-0.90546876){$M$}
\usefont{T1}{ptm}{m}{n}
\rput(0.3,-0.7054688){$s$}
\usefont{T1}{ptm}{m}{n}
\rput(4.55,-0.7054688){$t$}
\usefont{T1}{ptm}{m}{n}
\rput(2.5,-0.98){$v$}
\end{pspicture}
}
\end{center}
\end{example}
The general setting of the game is representable as a game 
in extensive form. 
However, there is no concept in game theory to construct an equilibrium
for an infinite game in extensive form
(only bargaining models such as the Rubinstein bargaining game are well studied,
cf.~\citet[ch.~16]{osb04}).
Thus, we have to impose reasonable conditions to guarantee finiteness of the game.

A straightforward idea would be to restrict the game to {\em simple paths}
and request that each vertex may be visited at most once.
While this restriction would lead to a simpler setting of the game,
it would also rule out very reasonable strategic behavior,
such as going through a cycle in the graph.
Indeed, a cycle of even length can not be seen as a reasonable choice
for any player since it necessarily increases the total costs for both players.
However, in the above Example~\ref{ex:infinite} it would be perfectly reasonable
for $B$ to enter the cycle of {\em odd} length and thus switch the role of the decider in $v$.
Therefore, a second visit of a vertex may well make sense.
However, if also $A$ enters the cycle in the next visit of $v$,
two rounds through the odd cycle constitute a cycle of even length which
we rejected before.
Based on these arguments we will impose the following restriction
which permits a rather general setting of the game:
\begin{quote}
\textbf{(R2)} The players can not select an arc which implies necessarily
a cycle of even length.
\end{quote}
Note that (R2) implies that an odd cycle may be part of the solution
path, but it may not be traversed twice (thus reversing the switching
between the players) since this would constitute a cycle of even length.
This aspect of allowing odd cycles will be taken explicitly into account
in the presented algorithms.
In the remainder of the paper we use ``cycle'' for any
closed walk, also if vertices are visited multiple times.
It also follows that each player can decide in each vertex at most once
and any arc can be used at most once by each player.

A different approach to guarantee finiteness of the game would
be to impose an upper bound on the total cost accrued by each player.
 This may seem relevant especially for answering the decision version
 where we ask whether the costs 
for both players remain
 below given bounds (see Section~\ref{sec:optspgd}).
Clearly, such an upper bound 
permits a clear answer to problems such as Example~\ref{ex:infinite}.
In that case, the precise value of the bound defines which of the two
players has to pay $M$ after a possibly large number of rounds through the cycle.
Anticipating this outcome, 
the unfortunate player will accept the cost $M$ the first time it has to decide in $v$.
However, it should be pointed out that a slight change in the upper bound
may completely turn around the situation and cause the other player to pay $M$.
Thus, the outcome of the game would depend on the remainder of the division of the
bound by the cycle cost.
This would cause a highly erratic solution structure
and does not permit a consistent answer to the decision problem.

\medskip
As an illustrative example of the game consider the following scenario. 
Two scientists $A$ and $B$ meet at a conference and lay out the plan for a joint paper.
This requires the completion of a number of tasks.
Each task requires a certain working time (identical for both $A$ and $B$)
to be completed.
Moreover, there is a precedence structure on the set of tasks,
e.g.\ the notation and definitions have to be finished
before the formal statement of results,
which in turn should precede the writing of the proofs etc.
Obviously, this implies a partial ordering on the set of tasks.
We can represent any state of the paper by a vertex
of a directed graph with an arc representing
a feasible task whose completion leads to a new state (=vertex). \newline
To avoid confusion and a mix-up of file versions 
the two scientists decide to alternate in working on the paper
such that each of them finishes a task and then passes the
file to the other co-author.
Since none of them wants to patronise the other they have 
no fixed plan on the division of tasks, 
but each can pick any outstanding task
as long as it is a feasible step to a new state of the paper.
They continue in this way until the paper is finished,
i.e.\ the final state is reached. 
Since both scientists are extremely busy each of them
tries to minimize the time devoted to the paper
taking into account the anticipated rational optimal
decisions by the other. \newline
More generally, the considered game applies to situations in which a system needs to be transformed from status 
$s$ to status $t$ by traversing several transitional status' in alternately performed steps.

\subsection{Solution Concept}
\label{sec:solcon}

In classical game theory \spg\ is a finite game in extensive form.
All feasible decisions for the players can be represented in a game tree,
where each node corresponds to the decision of a certain player in a
vertex of the graph $G$.

The standard procedure to determine equilibria in a game tree is {\em backward induction}
(see \citet[ch.~5]{osb04}).
This means that for each node in the game tree,
whose child nodes are all leaves, the associated player can reach
a decision by simply choosing the best of all child nodes w.r.t.\ their
allocated total cost, i.e.\ the cost of the corresponding path in $G$
attributed to the current player.
Then these leaf nodes can be deleted and the
pair of costs of the chosen leaf is moved to its parent node.
In this way, we can move upwards in the game tree towards the root and settle
all decisions along the way.

This backward induction procedure implies a strategy for each player,
i.e.\ a rule specifying for each node of the game tree
associated with this player
which arc to select in the corresponding vertex of $G$:
{\em Always choose the arc according to the result of backward induction.}
Such a strategy for both players is a {\em Nash equilibrium} and also a so-called
{\em subgame perfect equilibrium} (a slightly stronger property),
since the decisions made in the underlying backward induction procedure
are also optimal for every subtree.\footnote{In order to guarantee a unique solution
of such a game and thus a specific subgame perfect Nash equilibrium,
we have to define a tie-breaking rule.
We will use the ``optimistic case'', where in case of indifference
a player chooses the option with lowest possible cost for the other player.
If both players have the same cost, the corresponding paths in the graph
are completely equivalent.
Assigning arbitrary but fixed numbers to each vertex in the beginning,
e.g.\ $1,\ldots,n$, we choose the path to a vertex with lowest vertex number.
}

The outcome, if both players follow this strategy,
is a unique path from $s$ to $t$ in $G$ corresponding
to the unique {\bf s}ubgame {\bf p}erfect {\bf e}quilibrium (SPE) which we will call {\bf \opath}.
A \opath\ for \spg\ is the particular solution in the game tree
with minimal cost for both selfish players
under the assumption that they have complete and perfect information
of the game and know that the opponent will
also strive for its own selfish optimal value.

Clearly, such a \opath\ path can be computed in exponential time
by exploring the full game tree.
It is the main goal of this paper to
study the complexity status of finding this \opath.
In particular, we want to establish the hardness of computation for general graphs
and identify special graph classes where a \opath\ can be found
without exploring the exponential size game tree.

An illustration is given in Example~\ref{ex:strat}.
Note that in general game theory one considers only outcomes
of strategies and their payoffs,
i.e.\ costs of paths from $s$ to $t$ in our scenario.
In this paper we will consider in each node of the game tree
the cost for each player for moving from the corresponding vertex $v$ of $G$
towards $t$, since the cost of the path from $s$ to $v$ does not influence
the decision in $v$.
This allows us to solve identical subtrees that appear in multiple places
of the game tree only once and use the resulting optimal subpath
on all positions.

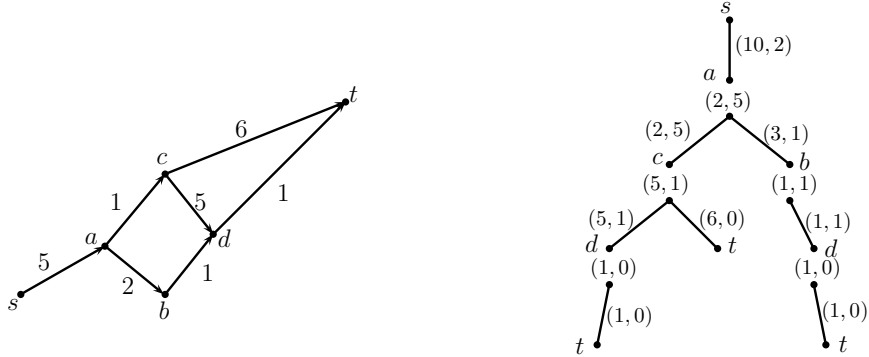
\begin{center}
\scalebox{0.8} 
{
\begin{pspicture}(0,-3.0595312)(14.8225,3.0595312)
\psdots[dotsize=0.12](1.8434376,-1.1548438)
\psdots[dotsize=0.12](2.8434374,0.04515625)
\psdots[dotsize=0.12](3.6434374,-0.95484376)
\psdots[dotsize=0.12](0.4434375,-1.9548438)
\psdots[dotsize=0.12](5.8434377,1.2451563)
\psline[linewidth=0.04cm,arrowsize=0.05291667cm 2.0,arrowlength=1.4,arrowinset=0.4]{->}(0.4434375,-1.9548438)(1.8434376,-1.1548438)
\psline[linewidth=0.04cm,arrowsize=0.05291667cm 2.0,arrowlength=1.4,arrowinset=0.4]{->}(1.8434376,-1.1548438)(2.8434374,0.04515625)
\psline[linewidth=0.04cm,arrowsize=0.05291667cm 2.0,arrowlength=1.4,arrowinset=0.4]{->}(2.8434374,0.04515625)(3.6434374,-0.95484376)
\psline[linewidth=0.04cm,arrowsize=0.05291667cm 2.0,arrowlength=1.4,arrowinset=0.4]{->}(3.6434374,-0.95484376)(5.8434377,1.2451563)
\usefont{T1}{ppl}{m}{n}
\rput(0.316875,-2.1198437){\large $s$}
\psdots[dotsize=0.12](2.8434374,-1.9548438)
\psline[linewidth=0.04cm,arrowsize=0.05291667cm 2.0,arrowlength=1.4,arrowinset=0.4]{->}(2.8434374,0.04515625)(5.8434377,1.2451563)
\usefont{T1}{ppl}{m}{n}
\rput(5.976875,1.3801563){\large $t$}
\usefont{T1}{ppl}{m}{n}
\rput(0.826875,-1.4198438){\large $5$}
\usefont{T1}{ppl}{m}{n}
\rput(2.226875,-1.8198438){\large $2$}
\usefont{T1}{ppl}{m}{n}
\rput(3.546875,-1.5998437){\large $1$}
\usefont{T1}{ppl}{m}{n}
\rput(4.806875,-0.25984374){\large $1$}
\usefont{T1}{ppl}{m}{n}
\rput(4.106875,0.80015624){\large $6$}
\usefont{T1}{ppl}{m}{n}
\rput(2.026875,-0.41984376){\large $1$}
\usefont{T1}{ppl}{m}{n}
\rput(3.426875,-0.41984376){\large $5$}
\usefont{T1}{ppl}{m}{n}
\rput(1.616875,-1.0198437){\large $a$}
\usefont{T1}{ppl}{m}{n}
\rput(2.826875,-2.2198439){\large $b$}
\usefont{T1}{ppl}{m}{n}
\rput(2.786875,0.30015624){\large $c$}
\usefont{T1}{ppl}{m}{n}
\rput(3.826875,-1.0198437){\large $d$}
\psdots[dotsize=0.12](12.223437,2.6051562)
\psline[linewidth=0.04cm](12.223437,2.6051562)(12.223437,1.6051563)
\psdots[dotsize=0.12](12.223437,1.6051563)
\usefont{T1}{ppl}{m}{n}
\rput(12.156875,2.8201563){\large $s$}
\usefont{T1}{ppl}{m}{n}
\rput(11.896875,1.7001562){\large $a$}
\psline[linewidth=0.04cm](12.223437,1.0051563)(11.223437,0.20515625)
\psline[linewidth=0.04cm](12.223437,1.0051563)(13.223437,0.20515625)
\psdots[dotsize=0.12](11.223437,-0.39484376)
\psdots[dotsize=0.12](13.223437,-0.39484376)
\usefont{T1}{ppl}{m}{n}
\rput(11.026875,0.30015624){\large $c$}
\usefont{T1}{ppl}{m}{n}
\rput(13.466875,0.26015624){\large $b$}
\psline[linewidth=0.04cm,arrowsize=0.05291667cm 2.0,arrowlength=1.4,arrowinset=0.4]{->}(2.8434374,-1.9548438)(3.6434374,-0.95484376)
\psline[linewidth=0.04cm](11.223437,-0.39484376)(10.223437,-1.1948438)
\psline[linewidth=0.04cm](11.223437,-0.39484376)(12.0234375,-1.1948438)
\psdots[dotsize=0.12](10.223437,-1.7948438)
\usefont{T1}{ppl}{m}{n}
\rput(9.926875,-1.0998437){\large $d$}
\usefont{T1}{ppl}{m}{n}
\rput(12.276875,-1.1398437){\large $t$}
\psline[linewidth=0.04cm](10.223437,-1.7948438)(10.0234375,-2.7948437)
\psdots[dotsize=0.12](10.0234375,-2.7948437)
\usefont{T1}{ppl}{m}{n}
\rput(9.736875,-2.8198438){\large $t$}
\usefont{T1}{ppl}{m}{n}
\rput(13.906875,-1.1798438){\large $d$}
\psline[linewidth=0.04cm](13.223437,-0.39484376)(13.623438,-1.1948438)
\psdots[dotsize=0.12](13.623438,-1.7948438)
\psline[linewidth=0.04cm](13.623438,-1.7948438)(13.823438,-2.7948437)
\psdots[dotsize=0.12](13.823438,-2.7948437)
\usefont{T1}{ppl}{m}{n}
\rput(14.116875,-2.7798438){\large $t$}
\usefont{T1}{ppl}{m}{n}
\rput(10.557969,-2.3248436){$(1,0)$}
\usefont{T1}{ppl}{m}{n}
\rput(14.137969,-2.2448437){$(1,0)$}
\usefont{T1}{ppl}{m}{n}
\rput(12.097969,-0.70484376){$(6,0)$}
\usefont{T1}{ppl}{m}{n}
\rput(10.257969,-0.70484376){$(5,1)$}
\usefont{T1}{ppl}{m}{n}
\rput(13.857968,-0.76484376){$(1,1)$}
\usefont{T1}{ppl}{m}{n}
\rput(10.297969,-1.5248437){$(1,0)$}
\usefont{T1}{ppl}{m}{n}
\rput(13.677969,-1.5248437){$(1,0)$}
\usefont{T1}{ppl}{m}{n}
\rput(11.1579685,-0.12484375){$(5,1)$}
\usefont{T1}{ppl}{m}{n}
\rput(13.297969,-0.12484375){$(1,1)$}
\psdots[dotsize=0.12](12.223437,1.0051563)
\psdots[dotsize=0.12](11.223437,0.20515625)
\psdots[dotsize=0.12](13.223437,0.20515625)
\psdots[dotsize=0.12](10.223437,-1.1948438)
\psdots[dotsize=0.12](12.0234375,-1.1948438)
\psdots[dotsize=0.12](13.623438,-1.1948438)
\usefont{T1}{ppl}{m}{n}
\rput(11.1979685,0.77515626){$(2,5)$}
\usefont{T1}{ppl}{m}{n}
\rput(13.1579685,0.6951563){$(3,1)$}
\usefont{T1}{ppl}{m}{n}
\rput(12.1979685,1.2551563){$(2,5)$}
\usefont{T1}{ppl}{m}{n}
\rput(12.787969,2.1751564){$(10,2)$}
\psline[linewidth=0.04cm,arrowsize=0.05291667cm 2.0,arrowlength=1.4,arrowinset=0.4]{->}(1.8234375,-1.1348437)(2.8234375,-1.9548438)
\end{pspicture}
}
\end{center}
\begin{example}\label{ex:strat}
Consider the above graph and the associated game tree.
The \opath\ is determined by backward induction and
represented by ordered pairs of cost values $(x,y)$
meaning that the decider in a given vertex has to pay a total value of $x$
whereas the opponent has to pay a value of $y$.

Note that if the game would be played cooperatively
the shortest path of value $9$ would give a lower total cost than
the optimal solution of $(10,2)$ in our case.
\end{example}

In this setting finding the \opath\ for the two players is not
an optimization problem as dealt with in combinatorial optimization
but rather the identification
of two sequences of decisions for the two players
fulfilling a certain property in the game tree.

Comparing the outcome of our game, i.e.\ the total cost of the \opath,
with the cost of the shortest path obtained by a cooperative strategy
we can consider the {\em Price of Anarchy} (cf.~\cite{nrt07})
defined by the ratio between these two values.
However, it is easy to see that the Price of Anarchy can become arbitrarily large,
e.g.\ in the following example, where the ratio is $\frac{1+M}{2}$.

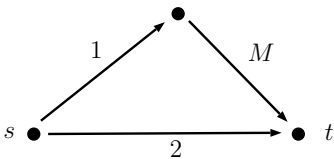
\begin{center}
\scalebox{0.8} 
{
\begin{pspicture}(0,-1.415)(5.46,1.3)
\psdots[linecolor=black, dotsize=0.22](0.5,-1.035)
\psdots[linecolor=black, dotsize=0.22](2.9,0.965)
\psdots[linecolor=black, dotsize=0.22](4.9,-1.035)
\psline[linecolor=black, linewidth=0.04, arrowsize=0.05291666666666667cm 2.0,arrowlength=1.4,arrowinset=0.0]{->}(0.74,-1.035)(4.62,-1.015)
\psline[linecolor=black, linewidth=0.04, arrowsize=0.05291666666666667cm 2.0,arrowlength=1.4,arrowinset=0.0]{->}(0.62,-0.815)(2.68,0.825)
\psline[linecolor=black, linewidth=0.04, arrowsize=0.05291666666666667cm 2.0,arrowlength=1.4,arrowinset=0.0]{->}(3.08,0.845)(4.72,-0.835)
\rput[bl](0.0,-1.095){\large{$s$}}
\rput[bl](5.34,-1.135){\large{$t$}}
\rput[bl](1.44,0.225){\large{$1$}}
\rput[bl](2.76,-1.415){\large{$2$}}
\rput[bl](4.06,0.185){\large{$M$}}
\end{pspicture}
}
\end{center}

\medskip

%
%

\subsection{Related literature}
\label{sec:lit}

A closely related game is known as \geo\ (see \citet{sch78}).
It is played on a directed graph with no costs.
Starting from a designated vertex $s\in V$, the two players move together
and take turns in selecting the next vertex.
The objective of the game is quite different from \spg,
namely, the game ends as soon as the players get stuck in a vertex
and the player who has no arc left for moving on loses the game.
Moreover, there is a further restriction that
in \geo\ each arc may be used at most once.

\citet{sch78} already showed \psp ness of \geo.
\citet{lisi80} proved that the variant \vgeo,
where each vertex cannot be visited more than once, is \psp\ for planar bipartite
graphs of bounded degree.
This was done as an intermediate step for showing that
\textsc{Go} is \psp.
\citet{frsi93} gave polynomial time algorithms for \geo\ and \vgeo\
when played on directed acyclic graphs.
In \citet{fsu93} it was proved that also the undirected variant of \geo\ is \psp.
However, if restricted to bipartite graphs they provided a polynomial time algorithm by using
linear algebraic methods on the bipartite adjacency matrix of the underlying graph.
Note that this result is in contrast to the \psp ness result of Section~\ref{sec:optspgu}
for \spg\ on bipartite undirected graphs.
\citet{bo93} showed that \vgeo\ is linear time solvable on both directed and 
undirected graphs of bounded treewidth. 
For \geo\ such a result was shown under
the additional restriction that the degree of every vertex is bounded by a constant 
- the unrestricted variant however is still open.

Recently, the \opath\ of \spg\ was used in~\citet{dkp13} as a criterion for
sharing the cost of the shortest path (in the classical sense) between two players.
A different variant of two players taking turns in the decision on a combinatorial
optimization problem and each of them optimizing its own objective function
was recently considered for the Subset Sum problem by \citet{dnp13}.

\subsection{Our contribution}
\label{sec:content}

We introduce the concept of \opath\ resulting from
backward induction in a game tree with complete and perfect information,
where two players pursue the optimization of their own
objective functions in a purely rational way.
Thus, a solution concept for the underlying game is determined which
incorporates in every step all anticipated decisions of future steps.

The main question we ask in this work concerns the complexity status of
computing such a \opath, if the game consists in the joint
exploration of a path from a source to a sink.
We believe that questions of this type could be an interesting
topic also for other problems on graphs and beyond.

We can show in Section~\ref{sec:sigmaspg} that for {\em directed graphs}
\spg\ is \psp\ even for bipartite graphs, while
for acyclic directed graphs a linear time algorithm 
follows from topological sort (Section~\ref{sec:acyclicspg}).
These results are in line with results from the literature
for the related game \geo.

On the other hand, for {\em undirected graphs}
we can show in Section~\ref {sec:optspgu} that again \spg\ is \psp\
even for bipartite graphs by a fairly complicated reduction
from \textsc{Quantified $3$-Sat}
while the related problem \geo\ is polynomially solvable
on undirected bipartite graphs.
This surprising difference shows that finding paths with minimal costs
can lead to dramatically harder problems than paths concerned only with reachability.

In Section~\ref{sec:cactus}
we give a fairly involved algorithm to determine the \opath\
on undirected {\em cactus graphs} in polynomial time.
It is based on several dynamic programming arrays
and a partitioning of the graph into components.
The running time of the algorithm can be bounded
by $O(n^2)$, where $n$ denotes the number of vertices in the graph.
The easier case of directed cactus graphs follows as a consequence.
We also argue that an extension of this partitioning technique to slightly more
general graphs, such as outerplanar graphs, is impossible.

A preliminary version describing some of the results given in this paper
but omitting most of the proofs
and only sketching the algorithms appeared as~\citet{dps14}.

\section{Spe-paths for \spg\ on directed graphs}
\label{sec:optspgd}

In general, there seems to be no way to avoid the exploration of the
exponential decision tree.
In fact, we can show a strong negative result.
Let us first define the following decision problem.
\begin{quote}
\spgd:\\
Given a weighted graph $G$ with dedicated vertices $s$, $t$
and two positive values $C_A$, $C_B$,
does the \opath\ 
yield costs $c(A) \leq C_A$ and $c(B) \leq C_B$ ?
\end{quote}
This means that we ask whether the unique subgame perfect equilibrium
of the associated extensive form game
fulfills the given cost bounds $C_A$ and $C_B$.

\subsection{$\sf PSPACE$-completeness}
\label{sec:sigmaspg}

The \psp ness of \spg\ on general graphs can be shown
by constructing an instance of the problem such that the
\opath\ decides the winner of \vgeo.
In fact, we can give an even stronger result with little additional effort.

 \begin{theorem}\label{theo1} \spgd~is $\sf PSPACE$-complete even
for bipartite directed graphs.
\end{theorem}

\proof
Inclusion in $\sf PSPACE$ can be shown easily by considering that the height
of the game tree is bounded by $2 |A|$.
Hence, we can determine the \opath\ in polynomial space by exploring the game tree
in a DFS-way.
In every node currently under consideration
we have to keep a list of decisions (i.e.\ neighboring vertices in the graph)
still remaining to be explored
and the cost of the currently preferred subpath among all the options starting in this node
that were already explored.
By the DFS-processing there are at most $2 |A|$ nodes on the path from the root
to the current node for which this information has to be kept.

We provide a simple reduction from \vgeo, which is known to be \psp~for
planar bipartite directed graphs  where the in-degree and the
out-degree of a vertex is bounded by two and the degree is bounded by three (\citet{lisi80}).
For a given instance of \vgeo\ we construct an instance of \spg, such that the
\opath\ path decides the winner of \vgeo:
Given the planar bipartite directed graph $G=(V,A)$ of \vgeo\ with starting
vertex $s$, we can two-color the vertices of $V$ because $G$ is bipartite.
For the two-coloring, we use the
colors red and green and color the vertices such that $s$ is a green vertex.

Starting from a copy of $G$ we create a new graph $H$ for \spgd\ as follows:
First we assign a cost $1$ to every arc $e\in A$.
Let $M$ be a large number, e.g.\ $M:=|A|+1$.
Then we introduce a new vertex $t$ which we color
red, and an arc of weight $M$ from each green vertex to $t$.
Next, introduce a green vertex $z$ and an arc of weight $M$
from each red vertex to $z$.
Finally, introduce an arc of cost $1$ from $z$ to $t$.

The constants $C_A$ and $C_B$ are set to $C_A=2$ and $C_B = M$. This means that a ``yes''-instance
corresponds to player $A$ winning \vgeo.
It is not hard to see that $H$ is a  bipartite
directed graph. Note that since the constructed graph is bipartite the rule of
\vgeo\ saying that each arc can be used at most once is equivalent to
(R2) in \spg.

Whenever a player gets stuck in a vertex playing \vgeo,
it would be possible to continue the path in $H$ towards $t$
(possibly via $z$) by choosing the arc of cost $M$.
On the other hand, both players will avoid to use such
a costly arc as long as possible and only one such arc will be chosen.
Thus, the \opath\ for \spg\ will incur  $\leq 2$
cost to one player and exactly cost $M$ to the other player, who is
thereby identified as the loser of \vgeo.
This follows from the fact that
if both players follow the \opath, they can anticipate the loser.
If it is $A$, then this player will immediately go from $s$ to $t$.
If it is player $B$, then $A$ will choose an arc with cost
$1$, then $B$ will go to $z$ paying $M$, and $A$ from $z$ to $t$
at cost $1$.
\qed

Note that the result of Theorem~\ref{theo1} also follows from
Theorem~\ref{the:psp} be replacing each edge in the undirected graph
by two directed arcs.
However, we believe that the connection to \geo\ established
in the above proof is interesting in its own right.

{\bf Remark:}\\
The above theorem is true even under the following restrictions:
the indegree of the vertices is bounded by two, the outdegree is bounded
by three, and the degree of the vertices is bounded by four.
However the planarity is lost
and the method for getting a planar graph described in \citet{lisi80}
does not carry over to \spg\ in an obvious way.

\subsection{Directed acyclic graphs}

\label{sec:acyclicspg}

If the underlying graph $G$ is acyclic we can do better by devising
a strongly polynomial time dynamic programming algorithm. It is related
to a dynamic programming scheme for the longest path problem in acyclic
directed graphs.

For any directed acyclic graph we can determine in  $O(|A|)$ time
a {\em topological sort}, i.e.\ a linear ordering of the vertices,
such that all arcs point ``from left to right'' (see e.g.~\cite[ch.~22.4]{Cormen}).
W.l.o.g.\ we can assume that $s$ and $t$ are the first and last vertex in this ordering.





For each vertex $v\in V$ we define the following two dynamic programming arrays:\\
$p_{d}(v)$: minimal path cost to go from $v$ to $t$ for the player {\em deciding} in $v$.\\
$p_{f}(v)$: minimal path cost to go from $v$ to $t$ for the {\em follower},
i.e.\ the player \textbf{not} deciding in $v$.

For each vertex $v$ let $S(v):=\{u\mid(v,u)\in A\}$ be the set of successors of $v$.
After initializing $p_{d}(t)=p_{f}(t)=0$,
the algorithm goes through the remaining vertices in reverse order of the topological sort
and computes for every vertex $v$:

\begin{quote}

$u':=\arg\min\{c(v,u)+p_{f}(u)\mid u\in S(v)\}$\\
$p_{d}(v):=c(v,u')+p_{f}(u')$, \
$p_{f}(v):=p_{d}(u')$

\end{quote}

The correctness of the computation follows immediately from the
structure of the topological sort since all vertices in $S(v)$ are processed before $v$.
Without elaborating on a formal proof we state:

\begin{theorem}\label{th:timeacyc}
The \opath\ of \spg\ on acyclic directed graphs can be computed in $O(|A|)$ time. \end{theorem}

\smallskip

In Section~\ref{sec:cactus} we will consider another special graph class
and derive a polynomial time algorithm
for \spg\ on undirected cactus graphs.
We will argue that this result immediately yields a
polynomial time algorithm also for directed cactus graphs
as stated in Corollary~\ref{th:timedir}.

\section{Spe-Paths for \spg\ on undirected graphs}
\label{sec:optspgu}

We will provide a reduction from the following problem
\textsc{Quantified $3$-Sat }which is known to be $\sf PSPACE$-complete
(\citet{stocki}).

{\bf Definition} (\textsc{\noun{Quantified $3$-Sat}}):
\begin{quote}
GIVEN:  Set $X=\{x_{1},\ldots,x_{n}\}$ of variables and a quantified
Boolean formula
$$F=(\exists x_{1})(\forall x_{2})(\exists x_{3})\ldots(\forall x_{n})\:\phi(x_{1},\ldots,x_{n})$$
where $\phi$ is a propositional formula over $X$ in $3$-$\sf CNF$
(i.e., in conjunctive normal form with exactly three literals per
clause).

QUESTION:   Is $F$ true?
\end{quote}
Let $C_{1},\ldots,C_{m}$ denote the clauses that make
up $\phi$, i.e., $\phi(x_{1},\ldots,x_{n})=C_{1}\wedge C_{2}\wedge\ldots C_{m}$,
where each $C_{i}$, $1\leq i\leq m$, contains exactly three literals.
\textsc{Quantified $3$-Sat} can be interpreted as
the following game (cf.~\citet{goldi}): There are two players (the
existential- and the universal-player) moving alternately, starting
with the existential-player. The $i$-th move consists of assigning
a truth value (``true'' or ``false'') to variable $x_{i}$. After
$n$ moves, the existential-player wins if and only if the produced
assignment makes $\phi$ true.


\subsection{$\sf PSPACE$-completeness}
\label{sec:pspaceu}

The following result shows a notable difference between \spg\ and \geo,
since \citet{fsu93} showed that \geo\ is polynomially solvable on
undirected bipartite graphs
(while $\sf PSPACE$-complete on general undirected graphs).

\begin{theorem}\label{the:psp}\spgd\ on undirected graphs is $\sf PSPACE$-complete,
even for bipartite graphs.
\end{theorem}

\begin{figure}
\begin{center}
\scalebox{0.92}
{
\begin{pspicture}(0,-10.39806)(13.6,10.418058) \psdots[dotsize=0.12](5.691875,10.039621) \psdots[dotsize=0.12](5.691875,10.039621) \psline[linewidth=0.04cm](5.691875,10.039621)(4.691875,9.439621) \psline[linewidth=0.04cm](4.691875,9.439621)(4.691875,8.639621) \psline[linewidth=0.04cm](4.691875,8.639621)(5.691875,8.039621) \psdots[dotsize=0.12](4.691875,9.439621) \psdots[dotsize=0.12](4.691875,8.639621) \psdots[dotsize=0.12](5.691875,8.039621) \psdots[dotsize=0.12](5.691875,8.039621) \psline[linewidth=0.04cm](5.691875,8.039621)(6.691875,8.639621) \psline[linewidth=0.04cm](6.691875,8.639621)(6.691875,9.439621) \psline[linewidth=0.04cm](6.691875,9.439621)(5.691875,10.039621) \psdots[dotsize=0.12](6.691875,9.439621) \psdots[dotsize=0.12](6.691875,8.639621) \psline[linewidth=0.04cm](5.691875,8.039621)(5.691875,7.239621) \psdots[dotsize=0.12](5.691875,7.239621) \psline[linewidth=0.04cm](5.691875,7.239621)(5.691875,6.439621) \psdots[dotsize=0.12](5.691875,6.439621) \psdots[dotsize=0.12](5.691875,6.439621) \psdots[
dotsize=0.12](5.691875,6.439621) \psline[linewidth=0.04cm](5.691875,6.439621)(4.691875,5.839621) \psline[linewidth=0.04cm](4.691875,5.039621)(4.691875,4.239621) \psline[linewidth=0.04cm](4.691875,4.239621)(5.691875,3.639621) \psdots[dotsize=0.12](4.691875,5.839621) \psdots[dotsize=0.12](4.691875,4.239621) \psdots[dotsize=0.12](5.691875,3.639621) \psdots[dotsize=0.12](5.691875,3.639621) \psline[linewidth=0.04cm](5.691875,3.639621)(6.691875,4.239621) \psline[linewidth=0.04cm](6.691875,4.239621)(6.691875,5.039621) \psline[linewidth=0.04cm](6.691875,5.839621)(5.691875,6.439621) \psdots[dotsize=0.12](6.691875,5.839621) \psdots[dotsize=0.12](6.691875,4.239621) \psline[linewidth=0.04cm](5.691875,3.639621)(5.691875,2.839621) \psdots[dotsize=0.12](4.691875,5.039621) \psdots[dotsize=0.12](6.691875,5.039621) \psline[linewidth=0.04cm](4.691875,5.839621)(4.691875,5.039621) \psline[linewidth=0.04cm](6.691875,5.839621)(6.691875,5.039621) \psdots[dotsize=0.12](5.691875,2.839621) \psdots[dotsize=0.12](5.691875,2.839621)
\psline[linewidth=0.04cm](5.691875,2.839621)(4.691875,2.239621) \psline[linewidth=0.04cm](4.691875,2.239621)(4.691875,1.439621) \psline[linewidth=0.04cm](4.691875,1.439621)(5.691875,0.839621) \psdots[dotsize=0.12](4.691875,2.239621) \psdots[dotsize=0.12](4.691875,1.439621) \psdots[dotsize=0.12](5.691875,0.839621) \psdots[dotsize=0.12](5.691875,0.839621) \psline[linewidth=0.04cm](5.691875,0.839621)(6.691875,1.439621) \psline[linewidth=0.04cm](6.691875,1.439621)(6.691875,2.239621) \psline[linewidth=0.04cm](6.691875,2.239621)(5.691875,2.839621) \psdots[dotsize=0.12](6.691875,2.239621) \psdots[dotsize=0.12](6.691875,1.439621) \psdots[dotsize=0.12](5.691875,-5.760379) \psdots[dotsize=0.12](5.691875,-5.760379) \psdots[dotsize=0.12](5.691875,-5.760379) \psline[linewidth=0.04cm](5.691875,-1.360379)(4.691875,-1.960379) \psline[linewidth=0.04cm](4.691875,-1.960379)(4.691875,-2.760379) \psline[linewidth=0.04cm](4.691875,-3.560379)(5.691875,-4.160379) \psdots[dotsize=0.12](4.691875,-2.760379) \psdots[dotsize=0.12](4.691875,-2.760379) \psdots[dotsize=0.12](5.691875,-4.160379) \psdots[dotsize=0.12](5.691875,-4.160379) \psline[linewidth=0.04cm](5.691875,-4.160379)(6.691875,-3.560379) \psline[linewidth=0.04cm](6.691875,-3.560379)(6.691875,-2.760379) \psline[linewidth=0.04cm](6.691875,-1.960379)(5.691875,-1.360379) \psdots[dotsize=0.12](6.691875,-1.960379) \psdots[dotsize=0.12](6.691875,-3.560379) \psdots[dotsize=0.12](4.691875,-3.560379) \psdots[dotsize=0.12](6.691875,-2.760379) \psline[linewidth=0.04cm](4.691875,-2.760379)(4.691875,-3.560379) \psline[linewidth=0.04cm](6.691875,-1.960379)(6.691875,-2.760379) \psline[linewidth=0.04cm,linestyle=dotted,dotsep=0.16cm](5.691875,0.839621)(5.691875,-1.360379) \psdots[dotsize=0.12](4.691875,-1.960379) \psline[linewidth=0.04cm](5.691875,-4.160379)(5.691875,-4.960379) \psdots[dotsize=0.12](5.691875,-4.960379) \psline[linewidth=0.04cm](5.691875,-4.960379)(5.691875,-5.760379) \psdots[dotsize=0.12](5.691875,-1.360379) \psline[linewidth=0.04cm](5.691875,-5.760379)(0.691875,-7.560379)
\psline[linewidth=0.04cm](5.691875,-5.760379)(2.491875,-7.560379) \psline[linewidth=0.04cm](5.691875,-5.760379)(11.291875,-7.560379) \psline[linewidth=0.04cm,linestyle=dotted,dotsep=0.16cm](4.491875,-7.560379)(8.691875,-7.560379) \psdots[dotsize=0.12](0.691875,-7.560379) \psdots[dotsize=0.12](2.491875,-7.560379) \psdots[dotsize=0.12](11.291875,-7.560379) \psline[linewidth=0.04cm](4.691875,9.439621)(0.691875,-7.560379) \psline[linewidth=0.04cm](0.691875,-7.560379)(6.691875,2.239621) \psline[linewidth=0.04cm](4.691875,5.039621)(0.691875,-7.560379) \psline[linewidth=0.04cm](2.491875,-7.560379)(4.691875,2.239621) \psline[linewidth=0.04cm](2.491875,-7.560379)(4.691875,9.439621) \psline[linewidth=0.04cm](6.691875,-2.760379)(11.291875,-7.560379) \usefont{T1}{ppl}{m}{n} \rput(5.86,-4.520379){$1$} \psline[linewidth=0.04cm](2.491875,-7.560379)(5.691875,-9.560379) \psdots[dotsize=0.12](5.691875,-9.560379) \psline[linewidth=0.04cm](0.691875,-7.560379)(5.691875,-9.560379) \psline[linewidth=0.04cm](5.691875,-9.560379)
(11.291875,-7.560379) \psdots[dotsize=0.12](12.091875,3.439621) \psbezier[linewidth=0.04](12.091875,3.439621)(12.091875,2.639621)(13.445432,-6.9275675)(12.891875,-7.760379)(12.338319,-8.59319)(6.324502,-10.378058)(5.711875,-9.620379) \psdots[dotsize=0.12](12.951875,-7.600379) \usefont{T1}{ppl}{m}{n} \rput(12.431875,3.439621){\large $t$} \usefont{T1}{ppl}{m}{n} \rput(5.661875,10.3){\large $s$} \psline[linewidth=0.04cm](6.691875,9.439621)(10.491875,3.439621) \psline[linewidth=0.04cm](4.691875,5.039621)(10.491875,3.439621) \psline[linewidth=0.04cm](10.491875,3.439621)(4.691875,2.239621) \psline[linewidth=0.04cm](6.691875,5.039621)(10.491875,3.439621) \psline[linewidth=0.04cm](4.691875,9.439621)(10.491875,3.439621) \psline[linewidth=0.04cm](6.691875,2.239621)(10.491875,3.439621) \psdots[dotsize=0.12](4.691875,-1.960379) \psline[linewidth=0.04cm](4.691875,-2.760379)(10.491875,3.439621) \psline[linewidth=0.04cm](6.691875,-2.760379)(10.491875,3.439621) \psdots[dotsize=0.12](10.491875,3.439621) \psline[linewidth=0.04cm](10.491875,3.439621)(12.091875,3.439621) \usefont{T1}{ppl}{m}{n} \rput(5.001875,9.879621){$x_1$} \usefont{T1}{ppl}{m}{n} \rput(6.431875,9.859621){$\bar{x}_1$} \usefont{T1}{ppl}{m}{n} \rput(6.491875,6.219621){$\bar{x}_2$} \usefont{T1}{ppl}{m}{n} \rput(6.231875,2.799621){$\bar{x}_3$} \usefont{T1}{ppl}{m}{n} \rput(4.901875,6.199621){$x_2$} \usefont{T1}{ppl}{m}{n} \rput(5.141875,2.759621){$x_3$} \usefont{T1}{ppl}{m}{n} \rput(5.021875,-1.520379){$x_{n}$} \usefont{T1}{ppl}{m}{n} \rput(6.311875,-1.520379){$\bar{x}_{n}$} \usefont{T1}{ppl}{m}{n} \rput(1.991875,-0.520379){$3$} \usefont{T1}{ppl}{m}{n} \rput(2.711875,-0.540379){$3$} \usefont{T1}{ppl}{m}{n} \rput(3.151875,-0.520379){$3$} \usefont{T1}{ppl}{m}{n} \rput(3.911875,-0.520379){$3$} \usefont{T1}{ppl}{m}{n} \rput(4.711875,-0.520379){$3$} \usefont{T1}{ppl}{m}{n} \rput(8.311875,-4.200379){$3$} \usefont{T1}{ppl}{m}{n} \rput(8.131875,-0.920379){$2$} \usefont{T1}{ppl}{m}{n} \rput(8.131875,1.3396211){$2$} \usefont{T1}{ppl}{m}{n} \rput(8.071875,2.439621){$2$}
\usefont{T1}{ppl}{m}{n} \rput(8.051875,3.159621){$2$} \usefont{T1}{ppl}{m}{n} \rput(8.091875,3.899621){$2$} \usefont{T1}{ppl}{m}{n} \rput(8.131875,4.719621){$2$} \usefont{T1}{ppl}{m}{n} \rput(7.991875,7.999621){$2$} \usefont{T1}{ppl}{m}{n} \rput(8.011875,6.399621){$2$} \usefont{T1}{ppl}{m}{n} \rput(0.441875,-7.800379){\large $c_1$} \usefont{T1}{ppl}{m}{n} \rput(2.381875,-7.820379){\large $c_2$} \usefont{T1}{ppl}{m}{n} \rput(11.381875,-7.780379){\large $c_m$} \usefont{T1}{ppl}{m}{n} \rput(5.651875,-9.860379){\large $w$} \usefont{T1}{ppl}{m}{n} \rput(13.241875,-7.620379){\large $d$} \usefont{T1}{ppl}{m}{n} \rput(10.671875,3.639621){\large $r$} \usefont{T1}{ppl}{m}{n} \rput(2.771875,-8.640379){$4$} \usefont{T1}{ppl}{m}{n} \rput(4.651875,-8.680379){$4$} \usefont{T1}{ppl}{m}{n} \rput(7.691875,-8.680379){$4$} \usefont{T1}{ppl}{m}{n} \rput(5.921875,-4.980379){\large $p$} \usefont{T1}{ppl}{m}{n} \rput(5.941875,-5.660379){\large $q$} \usefont{T1}{ppl}{m}{n} \rput(6.011875,7.239621){$u_2$} \usefont{T1}{ppl}
{m}{n} \rput(4.341875,9.419621){$v_{1,1}$} \usefont{T1}{ppl}{m}{n} \rput(4.121875,8.639621){$v_{1,2}$} \usefont{T1}{ppl}{m}{n} \rput(5.301875,7.999621){$v_{1,3}$} \usefont{T1}{ppl}{m}{n} \rput(6.961875,8.499621){$v_{1,4}$} \usefont{T1}{ppl}{m}{n} \rput(7.081875,9.459621){$v_{1,5}$} \usefont{T1}{ppl}{m}{n} \rput(5.321875,-4.240379){$v_{n,4}$} \psline[linewidth=0.04cm](4.691875,-2.760379)(2.491875,-7.560379) \usefont{T1}{ppl}{m}{n} \rput(3.791875,-4.260379){$3$}
\end{pspicture}  }
\end{center}
\vspace*{-8mm}
\caption{\label{fig:Undirected-graph-VarI}
Undirected graph $G$ of instance $\mathcal{S}$
constructed for an instance $\mathcal{Q}$ of \textsc{\noun{Quantified $3$-Sat}}
with $C_{1}=(\bar{x}_{1}\vee\bar{x}_{2}\vee x_{3})$ and $C_{2}=(\bar{x}_{1}\vee\bar{x}_{3}\vee\bar{x}_{n})$.}
\end{figure}
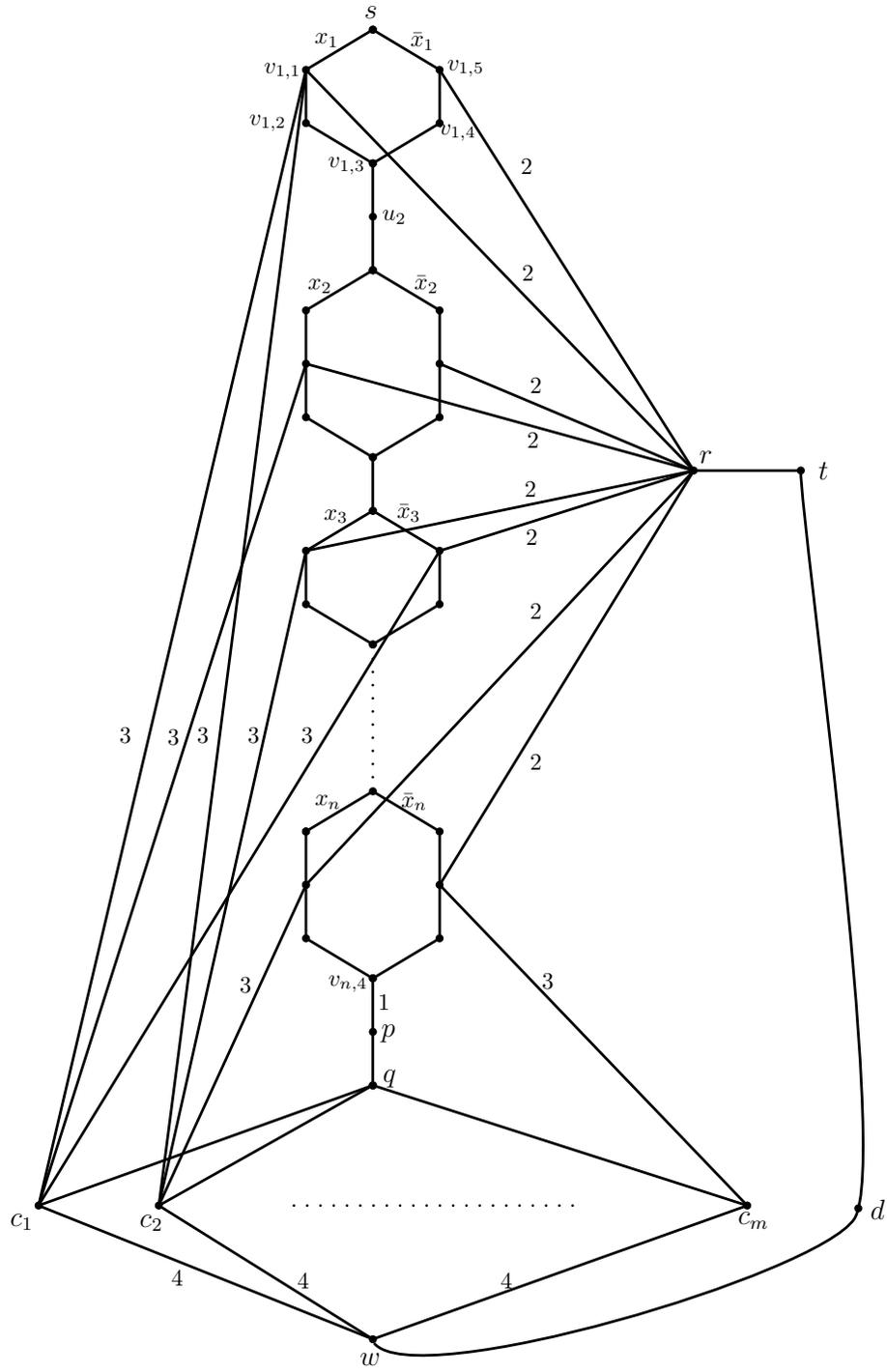

\proof
Inclusion in $\sf PSPACE$ follows from a similar argument as in the proof of Theorem~\ref{theo1}.
Given an instance $\mathcal{Q}$ of \textsc{\noun{Quantified $3$-Sat}}
we construct an instance $\mathcal{S}$ of \spg\
by creating an undirected graph $G=(V,E)$ as follows.
The vertices are
$2$-colored (using the colors red and green) to show that $G$ is
bipartite. To construct $G$, we introduce (see Figure~\ref{fig:Undirected-graph-VarI}):
\begin{itemize}
\setlength{\itemsep}{0pt}
\item green vertices $d,p,r$, red vertices $w,q,t$
\item edges $\{p,q\}$, $\{r,t\}$, $\{w,d\}$ and $\{d,t\}$
\item for each clause $C_{j}$, a green vertex $c_{j}$
\item for each even $i$, $2\leq i\leq n$, an ``octagon'', i.e.,
\begin{itemize}
\item red vertices $v_{i,0},v_{i,2},v_{i,4},v_{i,6}$
\item green vertices $v_{i,1},v_{i,3},v_{i,5},v_{i,7}$
\item edges $\{v_{i,\ell},v_{i,\ell+1}\}$, $0\leq\ell\leq6$, and edge
$\{v_{i,7},v_{i,0}\}$
\end{itemize}
\item for each odd $i$, $1\leq i\leq n$, a ``hexagon'', i.e.,
\begin{itemize}
\setlength{\itemsep}{0pt}
\item green vertices $v_{i,0},v_{i,2},v_{i,4}$
\item red vertices $v_{i,1},v_{i,3},v_{i,5}$
\item edges $\{v_{i,\ell},v_{i,\ell+1}\}$, $0\leq\ell\leq4$, and edge
$\{v_{i,5},v_{i,0}\}$
\end{itemize}
\end{itemize}
In order to connect these parts, we introduce:
\begin{itemize}
\item for each even $i$, $2\leq i\leq n$
\begin{itemize}
\item a green vertex $u_{i}$
\item edges $\{v_{i-1,3},u_{i}\}$, $\{u_{i},v_{i,0}\}$ and the edges $\{v_{i,2},r\}$,
$\{v_{i,6},r\}$
\item edge $\{v_{i,4},v_{i+1,0}\}$, where $v_{n+1,0}:=p$
\item for each clause $C_{j}$, the edge $\{v_{i,2},c_{j}\}$ if $x_{i}\in C_{j}$
and $\{v_{i,6},c_{j}\}$ if $\bar{x}_{i}\in C_{j}$
\end{itemize}
\item for each odd $i$, $1\leq i\leq n$
\begin{itemize}
\item the edges $\{v_{i,1},r\}$, $\{v_{i,5},r\}$
\item for each clause $C_{j}$, the edge $\{v_{i,1},c_{j}\}$ if $x_{i}\in C_{j}$
and $\{v_{i,5},c_{j}\}$ if $\bar{x}_{i}\in C_{j}$
\end{itemize}
\item for each $j$, $1\leq j\leq m$,
\begin{itemize}
\item edges $\{q,c_{j}\}$ and $\{w,c_{j}\}$
\end{itemize}
\end{itemize}

Abusing notation, for $1\leq i\leq n$, let $x_{i}:=\{v_{i,0},v_{i,1}\}$,
and $\bar{x}_{i}:=\{v_{i,0},v_{i,5}\}$ if $i$ is odd resp.\ $\bar{x}_{i}:=\{v_{i,0},v_{i,7}\}$
if $i$ is even, i.e., we identify a literal with an edge of the some label.
For illustration, in Figure~\ref{fig:Undirected-graph-VarI}
we assume $C_{1}=(\bar{x}_{1}\vee\bar{x}_{2}\vee x_{3})$ and $C_{2}=(\bar{x}_{1}\vee\bar{x}_{3}\vee\bar{x}_{n})$.

Finally, we define the edge costs.\footnote{In the introduction
edge costs were defined to be strictly positive.
For simplicity we use zero costs in this proof, but
these could be easily replaced by some small $\eps>0$.}
We start with the edges emanating
from vertex $w$: The cost of edge $\{w,d\}$ is $c\{w,d\}=0$, all
other edges emanating from $w$ have cost $4$. The edge $\{r,t\}$
has cost $c(\{r,t\})=0$, while each other edge emanating from $r$
has cost $2$. The edge $\{v_{n,4},p\}$ has cost $c(\{v_{n,4},p\})=1$.
For each $1\leq j\leq m$, the edges emanating from vertex $c_{j}$
which do not correspond to $\{c_{j},w\}$ or $\{c_{j}, q\}$ have cost
$3$. The remaining edges have zero cost. \\
Note that from the fact that each edge connects a green with a red
vertex, it immediately follows that $G$ is a bipartite graph. Now,
in $\mathcal{S}$ we set $C_A = 0$ and $C_B = 2$, and ask if the  \opath\ 
yields costs $c(A) \leq C_A$ and $c(B) \leq C_B$.

\textbf{Claim.} $\mathcal{Q}$ is a ``yes''-instance of \textsc{\noun{Quantified
$3$-Sat}} $\Leftrightarrow$  $\mathcal{S}$ is a ``yes''-instance of \spg.

\textbf{Proof of claim.} Player $A$ starts with the first move along
an edge emanating from $s$, i.e., $A$ moves along $x_{1}$ or $\bar{x}_{1}$.
Due to the fact that $s$ is a green vertex and $G$ is bipartite,
we can conclude that in a green vertex, $A$ needs to choose the next
edge, and in a red vertex, it is $B$'s turn to choose the next edge.
\\
W.l.o.g.\ assume $A$ moves along $x_{1}$ (which, as we will see
later, corresponds to setting to true $x_{1}$ in instance $\mathcal{Q}$).
Because of (R2), player $B$ cannot move back to $s$ along $x_{1}$.
Now, $B$ has three choices. $B$ either moves along (i) edge $\{v_{1,1},r\}$
with cost $2$, or (ii) edge $\{v_{1,1},v_{1,2}\}$ with zero cost,
or (iii) an edge of cost $3$ connecting $v_{1,1}$ with some vertex
$c_{j}$.\\
\uline{\noun{Case I:}}\emph{ }$B$ moves along \emph{$\{v_{1,1},r\}$}.\\
Then $A$ will use the edge $(r,t)$ of zero cost in order to minimize
her own total cost and \emph{the game ends with $c(A)=0$ and $c(B)=2$.}

As a consequence of the outcome in Case~I, $B$ will not choose (iii),
because $B$ would end with $c(B)\geq3$ in that case. Note that each
vertex $v_{i,1}$ for odd $i$ (resp.\ $v_{i,2}$ for even $i$)
is a red vertex, i.e., an analogous situation applies in each such
vertex. Thus, we can observe the following:
\begin{quote}
\textbf{(Observation~1)} Player $B$ does not move along an edge
of cost $3$, i.e., an edge $\{v_{i,1},c_{j}\}$ for odd $i$ resp.\
$\{v_{i,2},c_{j}\}$ for even $i$.
\end{quote}
\uline{\noun{Case II:}}\emph{ }$B$ moves along \emph{$\{v_{1,1},v_{1,2}\}$}.\\
In $v_{1,2}$, player $A$ needs to move along $\{v_{1,2},v_{1,3}\}$
(again, $A$ cannot move back to $v_{1,2}$ along $\{v_{1,1},v_{1,2}\}$
due to (R2)). At $v_{1,3}$, it is again player $B$'s turn. \\
\uline{\noun{Case }}\uline{IIa}\uline{\noun{:}}
$B$ moves along $\{v_{1,3},v_{1,4}\}$.\\
Then $A$ necessarily moves
along $\{v_{1,4},v_{1,5}\}$, leaving $B$ with the decision in $v_{1,5}$.
$B$ must not move along $\bar{x}_{1}$ because of (R2). Because of
Observation~1, this means that $B$ moves along $\{v_{1,5},r\}$
of cost $2$. Clearly, $A$ then chooses edge $(r,t)$, and again
\emph{the game ends with $c(A)=0$ and $c(B)=2$}. \\
\uline{\noun{Case }}\uline{IIb}\uline{\noun{:}}
$B$ moves along $\{v_{1,3},u_{2}\}$.\\
In the next step, $A$ moves to $v_{2,0}$. Now, it is $B$'s turn to pick
$x_{2}$ or $\bar{x}_{2}$, i.e., to decide which literal she sets
to true in instance $\mathcal{Q}$. W.l.o.g.\ assume $B$ moves along
$\bar{x}_{2}$. In the next step $A$ has only one edge to move along,
leaving $B$ with the decision in $v_{2,6}$. Analogously to above,
\emph{the game ends with $c(A)=0$ and $c(B)=2$} if $B$ moves
along $\{v_{2,6},r\}$. If $B$ does not move along $\{v_{2,6},r\}$,
the players continue until $v_{2,4}$ is reached, because $B$ does
not choose an edge of cost $3$ (see Observation~1). Again, it is
$B$'s turn:\\
If $B$ moves along $\{v_{2,4},v_{2,3}\}$, $A$ necessarily moves
along $\{v_{2,3},v_{2,2}\}$. In $v_{2,2}$, $B$ is not allowed to
move along $\{v_{2,2},v_{2,1}\}$ because this would result in a deadlock
($A$ would be unable to move along another edge due to (R1)). With
Observation~1, this means that $B$ moves along $\{v_{2,2},r\}$;
i.e., again \emph{the game ends with $c(A)=0$ and $c(B)=2$}. \\
Assume that $B$ moves along $\{v_{2,4},v_{3,0}\}$, leaving $A$ to choose
in $v_{3,0}$ between $x_{3}$ and $\bar{x}_{3}$.
It is easy to see that in what follows, repeatedly analogous decisions need to be
made. As a consequence, we can observe the following.
\begin{quote}
\textbf{(Observation~2)} If \spg~ends before vertex $v_{n,4}$
is reached, it ends with $c(A)=0$ and $c(B)=2$.

\textbf{(Observation~3)} If vertex $v_{n,4}$ is visited, then for
even $i$ player $B$ has chosen between $x_{i}$ and $\bar{x}_{i}$,
while for odd $i$ player $B$ has chosen between $x_{i}$ and $\bar{x}_{i}$.
\end{quote}
Assume that vertex $v_{n,4}$ is visited, where it is $B$'s turn
to take the next decision. Analogously to above we know that, if $B$
does not move along $\{v_{n,4},p\}$, then again the game ends with
$c(A)=0$ and $c(B)=2$. Assume $B$ moves along $\{v_{n,4},p\}$
with cost $1$. Clearly, in the next step $A$ has to move along
$\{p,q\}$ leaving $B$ with the decision in vertex $q$, i.e., $B$
has to choose a vertex $c_{j}$ (i.e, a clause $C_{j}$) to move to.
\\
Let $\phi$ be the truth assignment that sets to true exactly the
literals (edges) chosen by a player so far. We will argue that $B$
aims at picking a clause $C_{j}$ which is not satisfied by $\phi$.
We will illustrate this by assuming that $B$ decides to move to vertex
$c_{1}$ representing clause $C_{1}=(\bar{x}_{1}\vee\bar{x}_{2}\vee x_{3})$.
\\
If $C_{1}$ is not satisfied by $\phi$, i.e., if none of the literals
$\bar{x}_{1},\bar{x}_{2},x_{3}$ has been moved along, then all edges
$x_{1},x_{2},\bar{x}_{3}$ have been used already. As a result, $A$
is not allowed to move along the edges $\{c_{1},v_{1,1}\}$, $\{c_{1},v_{2,2}\}$
and $\{c_{1},v_{3,5}\}$ because of (R2). Again due to (R2), $A$
must not move back to $q$ along edge $\{c_{1},q\}$. Thus, $A$ needs
to move along $\{c_{1},w\}$ imposing a cost of $4$ on player $A$.
In the next steps, $B$ obviously moves along edge $\{w,d\}$, implying
that $A$ moves along $\{d,t\}$. \emph{Thus, the game ends with $c(A)=4$
and $c(B)=1$.}\\
On the other hand, if at least one of the literals $\bar{x}_{1},\bar{x}_{2},x_{3}$
have been used so far, then at least one of the edges $x_{1},x_{2},\bar{x}_{3}$
have not already been used. As a consequence, $A$ -- trying to avoid
the expensive edge $\{c_{1},w\}$ -- is able to use one of the edges
$\{c_{1},v_{1,1}\}$, $\{c_{1},v_{2,2}\}$ and $\{c_{1},v_{3,5}\}$.
W.l.o.g.\ assume that $\bar{x}_{3}$ has not been used and that $A$
now moves along $\{c_{1},v_{3,5}\}$ with cost $3$. At $v_{3,5}$,
$B$ cannot move along $\bar{x}_{3}$ due to (R2). If $B$ moves along
$\{v_{3,5},v_{3,4}\}$, a deadlock arises: $A$ is neither able to
move back to $v_{3,5}$ along $\{v_{3,5},v_{3,4}\}$, nor to move
along $\{v_{3,4},v_{3,3}\}$, because this would violate (R2) since
both $v_{3,3}$ and $v_{3,5}$ have already been visited. Due to (R1),
player $B$ hence must not use $\{v_{3,5},v_{3,4}\}$. Consequently,
player $B$ possibly needs to choose between an edge of cost $2$
or edge $\{v_{3,5},r\}$ with cost $2$. Clearly, $B$ chooses the
latter edge, since in vertex $r$ player $A$ will move along $\{r,t\}$.
\emph{Thus, the game ends with $c(A)=3$ and $c(B)=1+2=3$.}

Summing up the two cases we have:\\
If  $\mathcal{Q}$ is a ``yes''-instance 
then it is better for $B$ to move along an edge towards $r$ 
before reaching $p$,
resulting in an outcome with $c(A)=0$ and $c(B)=2$.\\
If  $\mathcal{Q}$ is a ``no''-instance 
there is a clause which is not satisfied by $\phi$
and $B$ will move along $\{v_{n,4},p\}$
resulting in $c(A)=4$ and $c(B)=1$.
 \qed

\medskip

\textbf{Remark.} Using the reduction provided in the above proof,
it is not hard to show that \spgd~remains $\sf PSPACE$-complete
in bipartite undirected graphs, even if (R2) is relaxed in the way
that we only require an edge not to be used more than once,
but do not rule out cycles of any length.

\subsection{Undirected acyclic graphs (trees)}
\label{sec:treesu}

Complementing Section~\ref{sec:acyclicspg}
we observe that in undirected acyclic graphs, i.e., trees, \spg\ becomes trivial:
Since in a tree there is exactly one path between two dedicated vertices,
the two player have no choice but to follow the unique path from $s$
to $t$.
Taking a diversion to a side branch, the players would always have to go
back the same way they came implying a cycle of even length.

\section{\spg\ on undirected cactus graphs}
\label{sec:cactus}

Since \spg\ is trivial on undirected trees and rather easy to solve in polynomial time
on directed acyclic graphs we try to push the bar a bit higher and
find polynomial algorithms on more general graph classes.
In this section we will show that a cactus graph still allows a polynomial
solution of \spg.
This is mainly due to a decomposition structure which allows to solve components of the graph
to optimality independently from the solution in the remaining graph.
However, we will illustrate by an example in Section~\ref{sec:moregeneral}
that more general graphs, such as outerplanar graphs, which are a superset of cactus graphs,
do not allow such a decomposition of the solution structure.
This gives a certain indication that \spg\ might become computationally intractable
on slightly more general graphs.

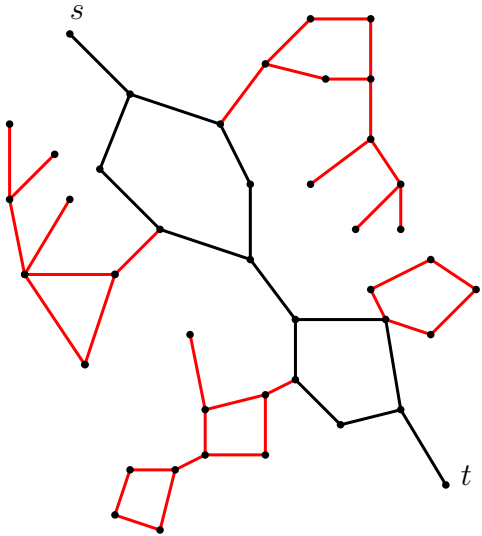
\begin{figure}[htb]
\begin{center}
{
\psscalebox{1.0 1.0} 
{
\begin{pspicture}(0,-3.5046394)(6.169279,3.5046394)
\definecolor{colour0}{rgb}{1.0,0.0,0.0}

\psline[linecolor=black, linewidth=0.04](0.84927887,3.1446393)(1.6492789,2.3446393)(1.2492789,1.3446393)(2.049279,0.5446393)(3.2492788,0.14463928)(3.2492788,1.1446393)(2.849279,1.9446393)(1.6492789,2.3446393)(1.6492789,2.3446393)
\psline[linecolor=black, linewidth=0.04](3.2492788,0.14463928)(3.849279,-0.6553607)(3.849279,-1.2553607)(3.849279,-1.4553608)(4.449279,-2.0553608)(5.249279,-1.8553607)(5.0492787,-0.6553607)(3.849279,-0.6553607)(3.849279,-0.6553607)
\psline[linecolor=black, linewidth=0.04](5.249279,-1.8553607)(5.849279,-2.8553607)(5.849279,-2.8553607)
\psline[linecolor=colour0, linewidth=0.04](2.049279,0.5446393)(1.4492788,-0.05536072)(0.24927887,-0.05536072)(1.0492789,-1.2553607)(1.4492788,-0.05536072)(1.4492788,-0.05536072)(1.4492788,-0.05536072)
\psline[linecolor=colour0, linewidth=0.04](2.849279,1.9446393)(3.4492788,2.7446394)(4.0492787,3.3446393)(4.849279,3.3446393)(4.849279,2.5446393)(4.249279,2.5446393)(3.4492788,2.7446394)(3.4492788,2.7446394)(3.4492788,2.7446394)
\psline[linecolor=colour0, linewidth=0.04](3.849279,-1.4553608)(3.4492788,-1.6553607)(2.6492789,-1.8553607)(2.6492789,-2.4553607)(3.4492788,-2.4553607)(3.4492788,-1.6553607)(3.4492788,-1.6553607)
\psline[linecolor=colour0, linewidth=0.04](2.6492789,-2.4553607)(2.2492788,-2.6553607)(1.6492789,-2.6553607)(1.4492788,-3.2553606)(2.049279,-3.4553607)(2.2492788,-2.6553607)(2.2492788,-2.6553607)
\psline[linecolor=colour0, linewidth=0.04](0.24927887,-0.05536072)(0.84927887,0.94463927)(0.84927887,0.94463927)
\psline[linecolor=colour0, linewidth=0.04](0.24927887,-0.05536072)(0.04927887,0.94463927)
\psline[linecolor=colour0, linewidth=0.04](0.04927887,0.94463927)(0.6492789,1.5446392)
\psline[linecolor=colour0, linewidth=0.04](0.04927887,0.94463927)(0.04927887,1.9446393)
\psline[linecolor=colour0, linewidth=0.04](4.849279,2.5446393)(4.849279,1.7446393)(4.0492787,1.1446393)(4.0492787,1.1446393)
\psline[linecolor=colour0, linewidth=0.04](4.849279,1.7446393)(5.249279,1.1446393)(4.6492786,0.5446393)(4.6492786,0.5446393)
\psline[linecolor=colour0, linewidth=0.04](5.249279,1.1446393)(5.249279,0.5446393)
\psline[linecolor=colour0, linewidth=0.04](2.6492789,-1.8553607)(2.4492788,-0.85536075)(2.4492788,-0.85536075)
\psline[linecolor=colour0,
linewidth=0.04](5.0492787,-0.6553607)(4.849279,-0.25536072)(5.6492786,0.14463928)(6.249279,-0.25536072)(5.6492786,-0.85536075)(5.0492787,-0.6553607)(5.0492787,-0.6553607)
\psdots[linecolor=black, dotsize=0.1](0.84927887,3.1446393)
\psdots[linecolor=black, dotsize=0.1](1.6492789,2.3446393)
\psdots[linecolor=black, dotsize=0.1](1.2492789,1.3446393)
\psdots[linecolor=black, dotsize=0.1](2.049279,0.5446393)
\psdots[linecolor=black, dotsize=0.1](3.2492788,1.1446393)
\psdots[linecolor=black, dotsize=0.1](2.849279,1.9446393)
\psdots[linecolor=black, dotsize=0.1](3.4492788,2.7446394)
\psdots[linecolor=black, dotsize=0.1](4.0492787,3.3446393)
\psdots[linecolor=black, dotsize=0.1](4.849279,3.3446393)
\psdots[linecolor=black, dotsize=0.1](4.849279,2.5446393)
\psdots[linecolor=black, dotsize=0.1](4.249279,2.5446393)
\psdots[linecolor=black, dotsize=0.1](3.2492788,0.14463928)
\psdots[linecolor=black, dotsize=0.1](3.849279,-0.6553607)
\psdots[linecolor=black, dotsize=0.1](3.849279,-1.4553608)
\psdots[linecolor=black, dotsize=0.1](4.449279,-2.0553608)
\psdots[linecolor=black, dotsize=0.1](5.249279,-1.8553607)
\psdots[linecolor=black, dotsize=0.1](5.0492787,-0.6553607)
\psdots[linecolor=black, dotsize=0.1](5.849279,-2.8553607)
\psdots[linecolor=black, dotsize=0.1](1.4492788,-0.05536072)
\psdots[linecolor=black, dotsize=0.1](0.24927887,-0.05536072)
\psdots[linecolor=black, dotsize=0.1](1.0492789,-1.2553607)
\psdots[linecolor=black, dotsize=0.1](3.4492788,-1.6553607)
\psdots[linecolor=black, dotsize=0.1](2.6492789,-1.8553607)
\psdots[linecolor=black, dotsize=0.1](2.6492789,-2.4553607)
\psdots[linecolor=black, dotsize=0.1](3.4492788,-2.4553607)
\psdots[linecolor=black, dotsize=0.1](2.2492788,-2.6553607)
\psdots[linecolor=black, dotsize=0.1](1.6492789,-2.6553607)
\psdots[linecolor=black, dotsize=0.1](1.4492788,-3.2553606)
\psdots[linecolor=black, dotsize=0.1](2.049279,-3.4553607)
\psdots[linecolor=black, dotsize=0.1](1.4492788,-0.05536072)
\psdots[linecolor=black, dotsize=0.1](1.0492789,-1.2553607)
\psdots[linecolor=black, dotsize=0.1](0.24927887,-0.05536072)
\psdots[linecolor=black, dotsize=0.1](0.04927887,0.94463927)
\psdots[linecolor=black, dotsize=0.1](0.6492789,1.5446392)
\psdots[linecolor=black, dotsize=0.1](0.04927887,1.9446393)
\psdots[linecolor=black, dotsize=0.1](0.84927887,0.94463927)
\psdots[linecolor=black, dotsize=0.1](4.849279,1.7446393)
\psdots[linecolor=black, dotsize=0.1](4.0492787,1.1446393)
\psdots[linecolor=black, dotsize=0.1](4.6492786,0.5446393)
\psdots[linecolor=black, dotsize=0.1](5.249279,1.1446393)
\psdots[linecolor=black, dotsize=0.1](5.249279,0.5446393)
\psdots[linecolor=black, dotsize=0.1](2.4492788,-0.85536075)
\rput[bl](0.84927887,3.3446393){\large $s$}
\rput[bl](6.0492787,-2.8553607){\large $t$}
\psdots[linecolor=black, dotsize=0.1](4.849279,-0.25536072)
\psdots[linecolor=black, dotsize=0.1](5.6492786,0.14463928)
\psdots[linecolor=black, dotsize=0.1](5.6492786,-0.85536075)
\psdots[linecolor=black, dotsize=0.1](6.249279,-0.25536072)
\end{pspicture}
}
}
\end{center}
\caption{\label{fig:connection} Graph $G$ with connection strip $G'$ (in black) and branches
$G\setminus G'$ (in red).}
\end{figure}

\begin{figure}[htb]
\begin{center}
{
\scalebox{0.9} 
{
\begin{pspicture}(0,-3.2521393)(5.389279,3.2521393)
\definecolor{colour0}{rgb}{1,0.0,0.0}
\psline[linecolor=black, linewidth=0.04](0.04927888,2.7971392)(0.8492789,1.9971393)(0.44927892,0.9971393)(1.249279,0.19713928)(2.4492788,-0.20286074)(2.4492788,0.7971393)(2.049279,1.5971392)(0.8492789,1.9971393)
\psline[linecolor=black, linewidth=0.04](2.4492788,-0.20286074)(3.049279,-1.0028607)(3.049279,-1.6028607)(3.049279,-1.8028609)(3.649279,-2.4028609)(4.449279,-2.2028608)(4.2492785,-1.0028607)(3.049279,-1.0028607)
\psline[linecolor=black, linewidth=0.04](4.449279,-2.2028608)(5.049279,-3.2028608)
\psdots[linecolor=black, dotsize=0.1](0.04927888,2.7971392)
\psdots[linecolor=black, dotsize=0.1](0.8492789,1.9971393)
\psdots[linecolor=black, dotsize=0.1](0.44927892,0.9971393)
\psdots[linecolor=black, dotsize=0.1](1.249279,0.19713928)
\psdots[linecolor=black, dotsize=0.1](2.4492788,0.7971393)
\psdots[linecolor=black, dotsize=0.1](2.049279,1.5971392)
\psdots[linecolor=black, dotsize=0.1](2.4492788,-0.20286074)
\psdots[linecolor=black, dotsize=0.1](3.049279,-1.0028607)
\psdots[linecolor=black, dotsize=0.1](3.049279,-1.8028609)
\psdots[linecolor=black, dotsize=0.1](3.649279,-2.4028609)
\psdots[linecolor=black, dotsize=0.1](4.449279,-2.2028608)
\psdots[linecolor=black, dotsize=0.1](4.2492785,-1.0028607)
\psdots[linecolor=black, dotsize=0.1](5.049279,-3.2028608)
\rput[bl](0.04927888,3.0621393){\large{$s$}}
\rput[bl](5.2492785,-3.1378608){\large{$t$}}
\psarc[linecolor=colour0, linewidth=0.04, dimen=outer, arrowsize=0.05291666666666667cm 2.0,arrowlength=1.4,arrowinset=0.0]{<-}(2.8,-2.0219233){0.26}{70.0}{350.0}
\psarc[linecolor=colour0, linewidth=0.04, dimen=outer, arrowsize=0.05291666666666667cm 2.0,arrowlength=1.4,arrowinset=0.0]{<-}(1.1,-0.061923217){0.26}{90.0}{10.0}
\psarc[linecolor=colour0, linewidth=0.04, dimen=outer, arrowsize=0.05291666666666667cm 2.0,arrowlength=1.4,arrowinset=0.0]{<-}(4.4,-0.9){0.26}{270.0}{180.0}
\psarc[linecolor=colour0, linewidth=0.04, dimen=outer, arrowsize=0.05291666666666667cm 2.0,arrowlength=1.4,arrowinset=0.0]{<-}(2.3,1.8){0.26}{270.0}{180.0}
\end{pspicture}
}
}
\end{center}
\caption{\label{fig:connection2} Graph $G$ with connection strip $G'$:
branches $G\setminus G'$ are contracted into swaps.}
\end{figure}
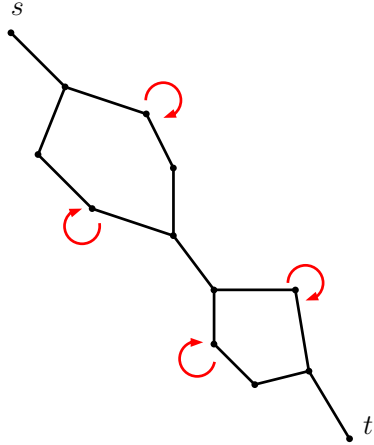

A {\em cactus graph} (also known as {\em Husimi tree}) is a graph
where each edge is contained in at most one simple cycle. 
Equivalently, any two simple cycles have at most one vertex in common.
This means that one could contract each cycle into a vertex in a unique way
and obtain a tree.
Note that cactus graphs are a subclass of series-parallel graphs and
thus have treewidth at most $2$.

Considering \spg, it is easy to see that the union of all simple paths from $s$ to $t$
define a subgraph $G'$ consisting of a unique sequence of edges (which are bridges of the graph)
and simple cycles.
We will call $G'$ the {\em connection strip} between $s$ and $t$.
All other vertices of the graph are ``dead end streets'', i.e.\ edges and cycles branching
off from $G'$ (see Figure~\ref{fig:connection}).
In the \opath\ vertices in $G\setminus G'$ could be included only to change the
role of the decision maker in a vertex of $G'$.
Clearly, any such deviation from $G'$ must be a cycle rooted in some vertex of $G'$.
Moreover, by (R2) only cycles (not necessarily simple) of odd length might be traversed in this way.
This structural property gives rise to a preprocessing step where all vertices in $G\setminus G'$
are contracted into a {\em swap option} in a vertex $v \in G'$
(see Figure~\ref{fig:connection2}) with cost $(sw_\d(v), sw_\f(v))$
meaning that if the path of the two players reaches a certain vertex $v \in G'$,
the current decider has the option to switch roles (by entering an odd cycle in $G\setminus G'$
rooted in $v$) at cost of $sw_\d(v)$ for himself (the decider) and $sw_\f(v)$ for the other player (the follower).

If there is more more than one cycle rooted in some vertex $v$,
each of them is contracted separately.
But since the \opath\ can utilize at most one swap in $v$,
we simply pick the swap option with the smallest cost for the decider.

Our algorithm will first compute these swap costs by recursively traversing the
components of $G\setminus G'$ in Section~\ref{sec:contr}.
Then, in the second step, the \opath\ in $G'$ is computed by moving backwards
from $t$ towards $s$ in Section~\ref{sec:mainpath}.
In each iteration of this second step a cycle is considered,
where the part of the \opath\ from the ``exit'' of the cycle towards $t$
is already known (resp.\ only a small number of possibilities remain).
By evaluating several dynamic programming arrays, the best options
for moving from the ``entry'' of this cycle to the exit are computed.
These iterations are continued until the starting vertex $s$ is reached.

\subsection{Contraction of the branches}
\label{sec:contr}

Consider a cycle $C(v_0)$ which is connected to the remaining graph only via $v_0$
and all other vertices of $C(v_0)$ have degree $2$, i.e.\ all other edges and cycles
incident to these vertices were contracted into swap options before.
For simplicity of notation we refer to vertices by their index number and
assume that $C(v_0)=C(0)$ consists of a sequence of vertices
$0,1,2,\ldots, k-1,k,0$.

\begin{figure}[htb]
\begin{center}
{
\scalebox{1.0} 
{
\begin{pspicture}(0,-2.51)(6.0378127,2.51)
\definecolor{colour1}{rgb}{1.0,0.0,0.0}
\psdots[dotsize=0.14](2.0,1.4)
\psdots[dotsize=0.14](1.0,0.8)
\psdots[dotsize=0.14](0.4,-0.6)
\psdots[dotsize=0.14](1.6,-1.4)
\psdots[dotsize=0.14](2.4,-2.2)
\psdots[dotsize=0.14](4.4,-1.8)
\psdots[dotsize=0.14](4.0,0.0)
\psdots[dotsize=0.14](3.4,1.4)
\psline[linewidth=0.04](2.0,2.4)(2.0,1.4)(1.0,0.8)(0.4,-0.6)(1.6,-1.4)(2.4,-2.2)(4.4,-1.8)(4.0,0.0)(3.4,1.4)(2.0,1.4)
\psline[linewidth=0.04,linecolor=colour1,arrowsize=0.05291667cm 2.0,arrowlength=1.4,arrowinset=0.0]{->}(1.6,2.5)(1.6,2.0)
\psline[linewidth=0.04,linecolor=colour1,arrowsize=0.05291667cm 2.0,arrowlength=1.4,arrowinset=0.0]{->}(2.2,2.0)(2.2,2.5)
\usefont{T1}{ptm}{m}{n}
\rput(2.5573437,1.14){\large $v_0=0$}
\usefont{T1}{ptm}{m}{n}
\rput(3.9473438,1.6677796){\large $k$}
\usefont{T1}{ptm}{m}{n}
\rput(4.7473435,0.10777956){\large $k-1$}
\usefont{T1}{ptm}{m}{n}
\rput(5.287344,-1.6322206){\large $\ldots$}
\usefont{T1}{ptm}{m}{n}
\rput(2.7773438,-1.7322205){\large $\l$}
\usefont{T1}{ptm}{m}{n}
\rput(2.1973438,-1.0522203){\large $\l-1$}
\usefont{T1}{ptm}{m}{n}
\rput(1.4873438,0.68777955){\large $1$}
\usefont{T1}{ptm}{m}{n}
\rput(1.3873438,-0.27222043){\large $\ldots$}
\usefont{T1}{ptm}{m}{n}
\rput(3.1573439,-0.11222046){\large $C(v_0)$}
\psline[linecolor=colour1, linewidth=0.04](1.6,2.11)(1.6,1.51)(0.8,0.91)(0.2,-0.69)(0.0,-0.69)(0.0,-0.49)(0.2,-0.49)(0.4,-0.89)(1.4,-1.69)(2.4,-2.49)(4.6,-1.89)(4.6,-1.69)(4.8,-1.69)(4.8,-1.89)(4.6,-1.89)(4.2,-0.09)(3.6,1.51)(2.2,1.51)(2.2,2.11)
\end{pspicture}
}
}
\end{center}
\caption{\label{fig:cycleround} Cycle $C(v_0)$ with a possible path (in red) going round the cycle (with possible swaps on the way).}
\end{figure}
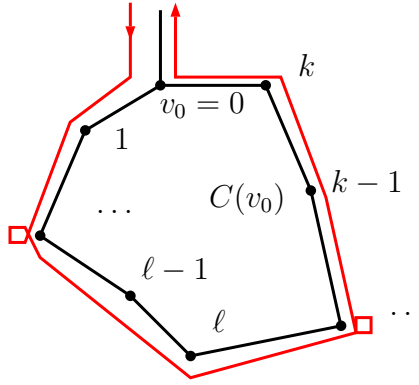

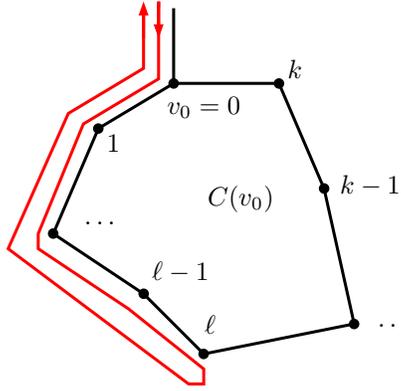
\begin{figure}[htb]
\begin{center}
{
\psscalebox{1.0 1.0} 
{
\begin{pspicture}(0,-3.11)(6.4689903,3.11)
\definecolor{colour1}{rgb}{1.0,0.0,0.0}
\psdots[linecolor=black, dotsize=0.14](4.0,0.91)
\psdots[linecolor=black, dotsize=0.14](3.0,0.31)
\psdots[linecolor=black, dotsize=0.14](2.4,-1.09)
\psdots[linecolor=black, dotsize=0.14](3.6,-1.89)
\psdots[linecolor=black, dotsize=0.14](4.4,-2.69)
\psdots[linecolor=black, dotsize=0.14](6.4,-2.29)
\psdots[linecolor=black, dotsize=0.14](6.0,-0.49)
\psdots[linecolor=black, dotsize=0.14](5.4,0.91)
\psline[linecolor=black, linewidth=0.04](4.0,1.91)(4.0,0.91)(3.0,0.31)(2.4,-1.09)(3.6,-1.89)(4.4,-2.69)(6.4,-2.29)(6.0,-0.49)(5.4,0.91)(4.0,0.91)(4.0,0.91)
\pscustom[linecolor=colour1, linewidth=0.04]
{
\newpath
\moveto(4.0,1.91)
}
\psline[linecolor=colour1, linewidth=0.04](3.8,1.91)(3.8,0.97)(2.8,0.37)(2.2,-1.09)(2.2,-1.29)(3.4,-2.09)(4.4,-2.89)(4.4,-3.09)(4.2,-3.09)(1.8,-1.29)(2.6,0.51)(3.6,1.11)(3.6,1.91)(3.6,1.91)
\psline[linecolor=colour1, linewidth=0.04, arrowsize=0.05291666666666667cm 2.0,arrowlength=1.4,arrowinset=0.0]{<-}(3.6,2)(3.6,1.51)(3.6,1.51)
\psline[linecolor=colour1, linewidth=0.04, arrowsize=0.05291666666666667cm 2.0,arrowlength=1.4,arrowinset=0.0]{<-}(3.8,1.51)(3.8,2)(3.8,1.91)

\rput[bl](3.8,0.45){ $v_0=0$}
\rput[bl](5.4,0.97777957){ $k$}
\rput[bl](6.1,-0.58222044){ $k-1$}
\rput[bl](6.6,-2.3222206){ $\ldots$}
\rput[bl](4.3,-2.4222205){ $\l$}
\rput[bl](3.6,-1.7422204){ $\l-1$}
\rput[bl](3.0,-0.002220459){ $1$}
\rput[bl](2.7,-0.96222043){ $\ldots$}
\rput[bl](4.34,-0.80222046){ $C(v_0)$}
\end{pspicture}
}
}
\end{center}
\caption{\label{fig:cycle} Cycle $C(v_0)$ with a possible path (in red) using a swap in vertex $\l$.}
\end{figure}

There are four possibilities how to use the cycle for a swap:
The players could enter the cycle by the edge $(0,1)$
and go around the full length of the cycle
(possibly using additional swaps in vertices of the cycle)
as depicted in Figure~\ref{fig:cycleround}.
Or after edge $(0,1)$ the players could move up to
some vertex $\l \in \{1,2,\ldots, k\}$,
turn around by utilizing a swap option in $\l$ and go back to $0$
the way they came as depicted in Figure~\ref{fig:cycle}.
Note that in the latter case, the players can not use
any additional swaps in vertices $1,\ldots, \l-1$
(resp.\ $k,k-1, \ldots, \l+1$)
since in that case the swap vertex would be visited three times
in violation of (R2).
Thus, we have to distinguish in each vertex whether such a turn around
is still possible or whether it is ruled out by a swap in
a previously visited vertex of the cycle.

These two configurations can also be used in a laterally reversed way
moving on the cycle in the different direction
starting with the edge $(0,k)$ which yields four cases in total.

\medskip
Let $D\in \{A,B\}$ be the decision maker in $0$.
We introduce the following generic notation for dynamic programming arrays:
\begin{quote}
$d^\pm_P(i)$: denotes the cost of a certain path starting in vertex $i$ and
ending in a fixed specified vertex.
\end{quote}
We use the subscript $P\in \{\d,\f\}$, where $P=\d$ signifies that the cost occurs for the player deciding in $i$
and $P=\f$ refers to the cost of the follower, i.e.\ the other player not deciding in $i$.
Superscript $\pm \in \{+,-\}$ shows that the decider in $i$ is equal to $D$
if $\pm=+$, or whether the other player decides in $i$, i.e.\ $\pm=-$.
For simplicity, we also extend the cost range and
use cost $\top$ if a path is infeasible.
When taking the minimum of values, $\top$ stands for an arbitrarily large value.
Adding the cost of a path to an array entry $\top$ yields again $\top$.

Following this system we define:\\
$tc^\pm_P(i):$ minimal cost to move from $i$ back to $0$, if a turn around is still possible.\\
$rc^\pm_P(i):$ minimal cost to move from $i$ back to $0$, if no turn around is possible
and the path has to go around the cycle, i.e.\ visit vertices $i+1, i+2, \ldots, k, 0$,
with possible swaps on the way.
If one player decides to turn around at some vertex $i$, the cost of the path back towards vertex $0$
is completely determined since no choices remain open.
The corresponding costs are independent from $D$ and will be recorded as $\mbox{path}_P(i)$
in analogy to above.

Now we can state the appropriate update recursion for the case where $D$ chooses
$(0,1)$ as a first edge.
We go backwards along this path and settle the minimal costs for vertices $k, k-1, \ldots, 1$.
The case where $D$ moves into the other direction of the cycle is completely analogous
and the final swap costs $sw(v_0)$ are given by the cheaper alternative.

\begin{figure}[htb]
\begin{center}
{
\psscalebox{0.9 0.9} 
{
\begin{pspicture}(0,-2.56)(7.22,2.56)
\psline[linecolor=black, linewidth=0.04](2.2,2.24)(1.2,0.64)(1.2,0.64)
\psline[linecolor=black, linewidth=0.04](2.2,2.24)(3.2,0.64)(3.2,0.64)
\rput[bl](1.6,2.3){player $A$}
\rput[bl](-0.3,0.24){$A$ moves to $i+1$}
\rput[bl](2.8,0.24){swap in $i$}
\rput[bl](2.8,-0.3){player $B$}
\psline[linecolor=black, linewidth=0.04](3.4,-0.36)(2.4,-1.76)(2.4,-1.76)
\psline[linecolor=black, linewidth=0.04](3.4,-0.36)(4.4,-1.76)
\rput[bl](0.4,-0.16){Case 1.}
\rput[bl](0.9,-2.1){$B$ moves to $i+1$}
\rput[bl](4.0,-2.1){$B$ turns and}
\rput[bl](4.0,-2.5){returns to $0$}
\rput[bl](4.2,-2.96){Case 2b.}
\rput[bl](1.6,-2.56){Case 2a.}
\end{pspicture}
}

}
\end{center}
\caption{\label{fig:dec} Decision tree for player $A$ deciding in vertex $i$.}
\end{figure}
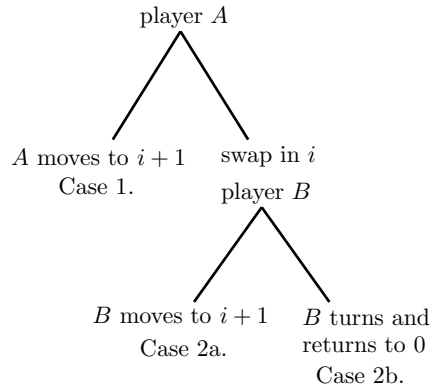

In any vertex $i$ the decision process has at most three outcomes
as illustrated in Figure~\ref{fig:dec}.
\begin{enumerate}
\item move on along the cycle to vertex $i+1$:

$rc^+_\d(i):=c(i,i+1)+rc^-_\f(i+1)$, $rc^+_\f(i)=rc^-_\d(i+1)$\\
$rc^-_\d(i):=c(i,i+1)+rc^+_\f(i+1)$, $rc^-_\f(i)=rc^+_\d(i+1)$\\
$tc^+_\d(i):=c(i,i+1)+tc^-_\f(i+1)$, $tc^+_\f(i)=tc^-_\d(i+1)$\\
$tc^-_\d(i):=c(i,i+1)+tc^+_\f(i+1)$, $tc^-_\f(i)=tc^+_\d(i+1)$

\item make a swap (if available):
Then the other player has (at most) two possibilities and chooses the one
with the lower cost between 2a.\ and 2b.\ (or the only feasible choice),
which automatically implies the cost for the decider in $i$.
\item[2a.] move on to vertex $i+1$:

$rc^+_\f(i)=sw_\f(i)+c(i,i+1)+rc^+_\f(i+1)$, $rc^+_\d(i):=sw_\d(i)+ rc^+_\d(i+1)$ \\
$rc^-_\f(i)=sw_\f(i)+c(i,i+1)+rc^-_\f(i+1)$, $rc^-_\d(i):=sw_\d(i)+ rc^-_\d(i+1)$ \\
$tc^+_\f(i):= sw_\f(i) + c(i,i+1)+ rc^+_\f(i+1)$, $tc^+_\d(i):= sw_\d(i) + rc^+_\d(i+1)$\\
$tc^-_\f(i):= sw_\f(i) + c(i,i+1)+ rc^-_\f(i+1)$, $tc^-_\d(i):= sw_\d(i) + rc^-_\d(i+1)$.

\item[2b.] turn around (if possible):
Since the decider at the end of the return path in vertex $0$ must
be different from $D$, the feasibility of a turn around depends on the number of
edges between $0$ and $i$.
\\
If $i$ is even:\\
$tc^+_\d(i):= sw_\d(i) + \mbox{path}_\f(i)$,
$tc^+_\f(i):= sw_\f(i) + \mbox{path}_\d(i)$\\
$tc^-_\d(i):= \top$, $tc^-_\f(i):= \top$

If $i$ is odd:\\
$tc^+_\d(i):= \top$, $tc^+_\f(i):= \top$\\
$tc^-_\d(i):= sw_\d(i) + \mbox{path}_\f(i)$,
$tc^-_\f(i):= sw_\f(i) + \mbox{path}_\d(i)$

\end{enumerate}
Now the decider in vertex $i$ can anticipate the potential decision of the other player
in case 2., since the other player will choose the better outcome between
cases 2a.\ and 2b.\ (if 2b.\ is feasible).
Hence, the decision maker in vertex $i$ chooses the minimum between
case 1.\ and case 2.\ (if a swap is possible)
independently for all four dynamic programming entries.
This immediately implies the cost for the other player.

It remains to discuss the initialization of the arrays for $i=k$, i.e.\ the last vertex in the cycle
and thus the first vertex considered in the recursion.
To avoid the repetition of all three cases implied by a possible swap at vertex $k$,
we add two artificial vertices $k+1$ and $k+2$ and three artificial edges $(k,k+1)$,
$(k+1,k+2)$ and $(k+2,0)$ replacing the previous edge $(k,0)$.
We set $c(k,k+1)=c(k,0)$ and the other two artificial edges have cost $0$.
It is easy to see that this extension of the cycle does not change anything.
Now we can start the recursive computation at vertex $k+2$ which has no swap option.
(If vertex $k$ does not have a swap option we can skip this extension and perform the following
initialization for $k$ replacing $k+2$.)
We get:

$rc^+_\d(k+2):=c(k+2,0)$, $rc^+_\f(k+2)=0$\\
$rc^-_\d(k+2):=\top$, $rc^-_\f(k+2)=\top$\\
$tc^+_\d(k+2):=c(k+2,0)$, $tc^+_\f(k+2)=0$\\
$tc^-_\d(k+2):=\top$, $tc^-_\f(k+2)=\top$

\subsection{Main part of the algorithm}
\label{sec:mainpath}

Now we proceed to the main part of the algorithm to determine the \opath\
for moving from $s$ to $t$ along the connection strip $G'$ after
the reminder of the graph was contracted into swap options in vertices of $G'$.
We traverse the connection strip backwards starting in $t$ and moving upwards in direction of $s$.
Recall that the connection strip consists of a sequence of cycles and edges.
In the following we will focus on the computation of an optimal subpath
for one cycle of this sequence.
Each such cycle has two designated vertices which all paths from $s$ to $t$ have to traverse,
an ``upper vertex'' $v_0$ through which every path starting in $s$ enters the cycle,
and a ``lower vertex'' $v_\l$ through which all paths connecting the cycle with $t$ have to leave the cycle.
As before we refer to the vertices of the cycle simply by their index numbers
and denote the cycle as $0, 1, \ldots, \l, \l+1, \ldots,  k, 0$.

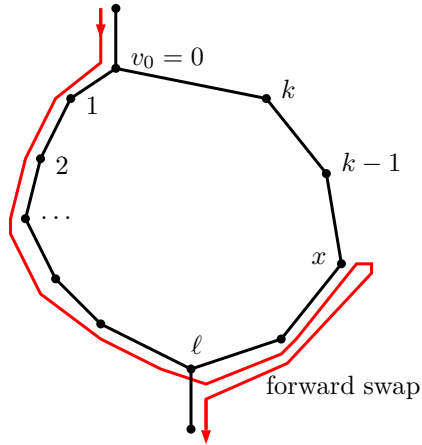
\begin{figure}[htb]
\begin{center}
{
\psscalebox{1.0 1.0} 
{
\begin{pspicture}(0,-2.8830466)(5.45,2.8830466)
\definecolor{colour0}{rgb}{1,0.0,0}
\psline[linecolor=black, linewidth=0.04](1.42,2.823912)(1.42,2.023912)(0.82,1.623912)(0.42,0.823912)(0.22,0.02391201)(0.62,-0.776088)(1.22,-1.376088)(2.42,-1.976088)(3.62,-1.576088)(4.42,-0.576088)(4.22,0.62391204)(3.42,1.623912)(1.42,2.023912)(1.42,2.023912)
\psdots[linecolor=black, dotsize=0.12](1.42,2.823912)
\psdots[linecolor=black, dotsize=0.12](1.42,2.023912)
\psdots[linecolor=black, dotsize=0.12](0.82,1.623912)
\psdots[linecolor=black, dotsize=0.12](0.42,0.823912)
\psdots[linecolor=black, dotsize=0.12](0.22,0.02391201)
\psdots[linecolor=black, dotsize=0.12](0.62,-0.776088)
\psdots[linecolor=black, dotsize=0.12](1.22,-1.376088)
\psdots[linecolor=black, dotsize=0.12](2.42,-1.976088)
\psdots[linecolor=black, dotsize=0.12](3.62,-1.576088)
\psdots[linecolor=black, dotsize=0.12](4.42,-0.576088)
\psdots[linecolor=black, dotsize=0.12](4.22,0.62391204)
\psdots[linecolor=black, dotsize=0.12](3.42,1.623912)
\psline[linecolor=colour0, linewidth=0.04](1.22,2.823912)(1.22,2.1)(0.62,1.623912)(0.22,0.823912)(0.02,0.02391201)
(0.02,-0.17608799)(0.42,-0.976088)(1.22,-1.576088)(2.02,-1.976088)(2.62,-2.176088)(3.62,-1.776088)
(3.82,-1.576088)(4.62,-0.576088)(4.82,-0.576088)(4.82,-0.7)(3.7,-1.9)(3.02,-2.2)
(2.62,-2.376088)(2.62,-2.776088)
\psline[linecolor=black, linewidth=0.04](2.42,-1.976088)(2.42,-2.776088)(2.42,-2.776088)
\psdots[linecolor=black, dotsize=0.12](2.42,-2.776088)
\psline[linecolor=colour0, linewidth=0.04, arrowsize=0.05291666666666668cm 2.0,arrowlength=1.4,arrowinset=0.0]{->}(1.22,2.823912)(1.22,2.423912)
\psline[linecolor=colour0, linewidth=0.04, arrowsize=0.05291666666666668cm 2.0,arrowlength=1.4,arrowinset=0.0]{->}(2.62,-2.376088)(2.62,-2.976088)
\rput[bl](1.62,2.023912){$v_0=0$}
\rput[bl](3.62,1.623912){$k$}
\rput[bl](4.42,0.62391204){$k-1$}
\rput[bl](4.02,-0.576088){$x$}
\rput[bl](1.02,1.423912){$1$}
\rput[bl](0.62,0.62391204){$2$}
\rput[bl](0.42,0.02391201){$\ldots$}
\rput[bl](2.42,-1.776088){$\ell$}
\rput[bl](3.42,-2.376088){forward swap}
\end{pspicture}
}
}
\end{center}
\caption{\label{fig:forward} Case (i): Forward swap in $x$.}
\end{figure}

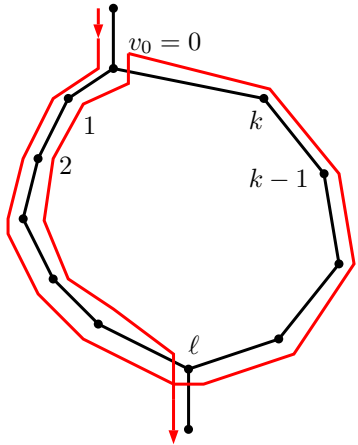
\begin{figure}[htb]
\begin{center}
{

\psscalebox{1.0 1.0} 
{
\begin{pspicture}(0,-2.8830466)(4.641033,2.8830466)
\definecolor{colour0}{rgb}{01,0.0,0}
\psline[linecolor=black, linewidth=0.04](1.42,2.823912)(1.42,2.023912)(0.82,1.623912)(0.42,0.823912)(0.22,0.02391201)(0.62,-0.776088)(1.22,-1.376088)(2.42,-1.976088)(3.62,-1.576088)(4.42,-0.576088)(4.22,0.62391204)(3.42,1.623912)(1.42,2.023912)(1.42,2.023912)
\psdots[linecolor=black, dotsize=0.12](1.42,2.823912)
\psdots[linecolor=black, dotsize=0.12](1.42,2.023912)
\psdots[linecolor=black, dotsize=0.12](0.82,1.623912)
\psdots[linecolor=black, dotsize=0.12](0.42,0.823912)
\psdots[linecolor=black, dotsize=0.12](0.22,0.02391201)
\psdots[linecolor=black, dotsize=0.12](0.62,-0.776088)
\psdots[linecolor=black, dotsize=0.12](1.22,-1.376088)
\psdots[linecolor=black, dotsize=0.12](2.42,-1.976088)
\psdots[linecolor=black, dotsize=0.12](3.62,-1.576088)
\psdots[linecolor=black, dotsize=0.12](4.42,-0.576088)
\psdots[linecolor=black, dotsize=0.12](4.22,0.62391204)
\psdots[linecolor=black, dotsize=0.12](3.42,1.623912)
\psline[linecolor=black, linewidth=0.04](2.42,-1.976088)(2.42,-2.776088)(2.42,-2.776088)
\psdots[linecolor=black, dotsize=0.12](2.42,-2.776088)
\psline[linecolor=colour0, linewidth=0.04, arrowsize=0.05291666666666668cm 2.0,arrowlength=1.4,arrowinset=0.0]{->}(1.22,2.823912)(1.22,2.423912)
\psline[linecolor=colour0, linewidth=0.04, arrowsize=0.05291666666666668cm 2.0,arrowlength=1.4,arrowinset=0.0]{->}(2.22,-2.376088)(2.22,-2.976088)
\rput[bl](1.62,2.223912){$v_0=0$}
\rput[bl](3.22,1.223912){$k$}
\rput[bl](3.22,0.42391202){$k-1$}
\rput[bl](1.02,1.15){$1$}
\rput[bl](0.7,0.62391204){$2$}
\rput[bl](2.42,-1.776088){$\ell$}
\psline[linecolor=colour0, linewidth=0.04](1.22,2.423912)(1.22,2.023912)(0.62,1.623912)(0.22,0.823912)(0.02,0.02391201)(0.02,-0.17608799)(0.42,-0.976088)(1.02,-1.576088)(2.22,-2.176088)(2.22,-2.176088)
\psline[linecolor=colour0, linewidth=0.04](2.22,-2.176088)(2.62,-2.176088)(3.82,-1.776088)(4.62,-0.576088)(4.42,0.62391204)(3.5,1.75)(1.62,2.223912)(1.62,2.223912)
\psline[linecolor=colour0, linewidth=0.04](1.62,2.223912)(1.62,1.823912)(1.02,1.55)(0.62,0.823912)(0.5,0.0)(0.82,-0.776088)(1.42,-1.176088)(2.22,-1.776088)(2.22,-2.376088)(2.22,-2.376088)
\end{pspicture}
}
}
\end{center}
\caption{\label{fig:forfull} Case (i), special case: Traverse the full cycle.}
\end{figure}

If we assume that the decision maker decides in $0$ to take the edge $(0,1)$
there are two main possibilities for the path from $0$ to $\l$ and onwards to the next cycle or edge.
The other situation, where the decision maker starts with the edge $(0,k)$ is
completely symmetric and the decider will finally take the better of the two options.

Case (i):
The two players may move along the vertices $0,1,\ldots, \l$, possibly with a few swaps on the way.
After reaching $\l$, they may either exit the cycle or continue to $\l+1, \ldots, x$,
make a {\em forward swap} in $x$ and return back via $x-1, x-2, \ldots $ back to $\l$ and finally exit the cycle (see Figure~\ref{fig:forward}).
As a special variant of this situation, the players may also never swap in some vertex $x$
but go back to $0$ thus traversing the full cycle and then taking the path $0,1,\ldots, \l$ a second time as depicted in Figure~\ref{fig:forfull}.

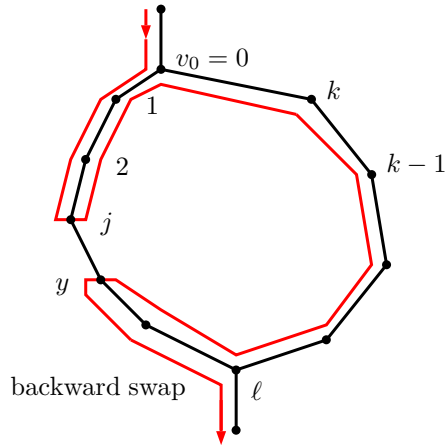
\begin{figure}[htb]
\begin{center}
{
\psscalebox{1.0 1.0} 
{
\begin{pspicture}(0,-2.8830466)(5.76,2.8830466)
\definecolor{colour0}{rgb}{1,0.0,0}
\psline[linecolor=black, linewidth=0.04](2.0,2.823912)(2.0,2.023912)(1.4,1.623912)(1.0,0.823912)(0.8,0.02391201)(1.2,-0.776088)(1.8,-1.376088)(3.0,-1.976088)(4.2,-1.576088)(5.0,-0.576088)(4.8,0.62391204)(4.0,1.623912)(2.0,2.023912)(2.0,2.023912)
\psline[linecolor=black, linewidth=0.04](3.0,-1.976088)(3.0,-2.776088)(3.0,-2.776088)
\psline[linecolor=colour0, linewidth=0.04, arrowsize=0.05291666666666668cm 2.0,arrowlength=1.4,arrowinset=0.0]{->}(1.8,2.823912)(1.8,2.423912)
\psline[linecolor=colour0, linewidth=0.04, arrowsize=0.05291666666666668cm 2.0,arrowlength=1.4,arrowinset=0.0]{->}(2.8,-2.376088)(2.8,-2.976088)
\rput[bl](2.2,2.023912){$v_0=0$}
\rput[bl](4.2,1.623912){$k$}
\rput[bl](5.0,0.62391204){$k-1$}
\rput[bl](1.8,1.423912){$1$}
\rput[bl](1.4,0.62391204){$2$}
\rput[bl](3.2,-2.376088){$\ell$}
\psline[linecolor=colour0, linewidth=0.04](1.8,2.423912)(1.8,2.223912)(1.8,2.023912)(1.2,1.623912)(0.8,0.823912)(0.6,0.02391201)(1.0,0.02391201)(1.2,0.823912)(1.6,1.623912)(2.0,1.823912)(3.8,1.423912)(3.8,1.423912)
\psline[linecolor=colour0, linewidth=0.04](3.8,1.423912)(4.6,0.62391204)(4.8,-0.576088)(4.2,-1.376088)(3.0,-1.776088)(2.0,-1.176088)(1.4,-0.776088)(1.0,-0.776088)(1.0,-0.976088)(1.6,-1.576088)(1.6,-1.576088)
\psline[linecolor=colour0, linewidth=0.04](1.6,-1.576088)(2.8,-2.176088)(2.8,-2.776088)(2.8,-2.776088)
\psdots[linecolor=black, dotsize=0.12](2.0,2.823912)
\psdots[linecolor=black, dotsize=0.12](2.0,2.023912)
\psdots[linecolor=black, dotsize=0.12](1.4,1.623912)
\psdots[linecolor=black, dotsize=0.12](1.0,0.823912)
\psdots[linecolor=black, dotsize=0.12](0.8,0.02391201)
\psdots[linecolor=black, dotsize=0.12](1.2,-0.776088)
\psdots[linecolor=black, dotsize=0.12](1.8,-1.376088)
\psdots[linecolor=black, dotsize=0.12](3.0,-1.976088)
\psdots[linecolor=black, dotsize=0.12](4.2,-1.576088)
\psdots[linecolor=black, dotsize=0.12](5.0,-0.576088)
\psdots[linecolor=black, dotsize=0.12](4.8,0.62391204)
\psdots[linecolor=black, dotsize=0.12](4.0,1.623912)
\psdots[linecolor=black, dotsize=0.12](3.0,-2.776088)
\rput[bl](1.2,-0.17608799){$j$}
\rput[bl](0.6,-0.976088){$y$}
\rput[bl](0.0,-2.376088){backward swap}
\end{pspicture}
}
}
\end{center}
\caption{\label{fig:backward} Case (ii): Backward swap in $y$.}
\end{figure}

Case (ii):
As a second, more complicated possibility, the two players may also move along vertices
$0,1, \ldots, j$, $j < \l$, and then utilize a swap in $j$ and return to $0$.
Then they are forced to move on from $0$ to $k, k-1, \ldots, \l$.
After reaching $\l$ they may either exit the cycle directly or they may also continue to
$\l-1, \l-2, \ldots, y$ with $y>j$, make a {\em backward swap} in $y$ and return via
$y +1, \ldots, \l$ where they finally exit the cycle (see Figure~\ref{fig:backward}).

Of course, all the resulting subpaths have to be feasible, in the sense that they actually
lead to the desired swap between decider and follower and contain no even cycles.
Moreover, all subpaths that are dead end streets used only for reaching a swap at some vertex
are traversed twice and thus are not allowed to contain any swap except at their endpoints.
In the above notation this holds for the subpaths $\l+1, \ldots, x$, the full cycle,
$0,\ldots, j$ and $\l, \ldots, y$.

\medskip
From our backward processing through the connection strip we can assume that we already know
the \opath\ from $v_\l$ to $t$ (without using any other vertex from the current cycle).
If there is a bridge leading from $v_\l$ towards $t$, it is easy to take the costs
of the \opath\ from $v_\l$ via this edge onwards to $t$ into account.
However, the situation gets more complicated if two cycles meet directly in one vertex,
i.e.\ if the lower vertex $v_\l$ of the current vertex is also the upper vertex of
the following cycle in the connection strip towards $t$.
In such a case we cannot combine the solutions of the two adjacent cycles in an arbitrary way
since each player can decide in vertex $v_\l$ at most once.

For the correct combination of solutions of two adjacent cycles
it will be necessary to distinguish whether
the path through the current cycle takes only one decision in $v_\l$ and then
leaves the cycle (Case (i) without forward swap in $x$ or
Case (ii) without backward swap in $y$)
or whether the first decision in $v_\l$ leads to another vertex
in the cycle and thus a second decision takes place in $v_\l$ with the only
possible result of leaving the cycle.
In the former case, we do not have any restriction on the solution of the following cycle
in the connection strip (closer to $t$)
since the path is also allowed to move back to $v_\l$
(which is the upper vertex of that cycle)
as it is done in Case (ii).
We denote by $\ed2v_P$ (exit costs) the costs of the \opath\ from $v_\l$ to $t$
allowing to visit $v_\l$ a second time.
In the latter case, we only consider paths that never move back to $v_\l$
in the following cycle (i.e.\ only Case (i) applies)
and denote the cost of the \opath\ from $v_\l$ to $t$ under this restriction by $\edv_P$.

\medskip
We now process the current cycle to determine the \opath\ starting in $v_0$.
Keeping the notational system introduced above we will introduce the following
dynamic programming arrays:

$td^\pm_P(i):$ minimal cost to move from $i$, $i=1,\ldots, \l-1$, to $\l$ and exit the cycle,
if a turn around (according to Case (i)) is still possible.\\
$rd^\pm_P(i):$ minimal cost to move from $i$, $i=1,\ldots, \l-1$, to $\l$ and exit the cycle,
if no turn around is possible.\\
$ad^\pm_P(i):$ (``alternative path'') minimal cost for moving from $i$,
$\l \leq i \leq k$, via $i-1, i-2,\ldots$ to $\l$ (Case (ii)).
We will later add a parameter $j$ to the definition of this array
since it may be combined with a backward swap (see below).\\
$fs^\pm_P(i):$ (``forward swap'') minimal cost for moving from $i$, $\l \leq i \leq k$,
without swaps to some vertex $x$, $i \leq x \leq k$, make a swap in $x$ and return to $\l$ (Case (i)).
The parity $\pm=+$ indicates that the decider in $i$ is the same as the decider in $\l$.
\\
$bs^\pm_P(i,j):$ (``backward swap'') minimal cost for moving from $i$, $j < i \leq \l$,
without swaps to some vertex $y$, $j < y \leq i$, make a swap in $y$ and return to $\l$ (Case (ii)).
Note that parameter $j$ indicates a lower bound on the position of the swap vertex $y$.
Since we do not know at this point at which vertex $j$ the path starting in $0$ was stopped by a swap, we have to compute the values of this array for all values of $j$, $j=1, \ldots, \l-1$.
The parity $\pm=+$ indicates that the decider in $i$ is the same as the decider in $\l$.

We first show how to compute the entries of the auxiliary dynamic programming arrays
starting with the forward swap moving backwards from a starting point $i=k'-1$ (see below)
down to $i=\l$.
Let $\mbox{fpath}_P(i)$ denote the cost of the path from $i$ back towards $\l$.
We have to take the minimum between the following two cases (if the second case exists):
\begin{enumerate}
\item move on along the cycle to vertex $i+1$:

$fs^+_\d(i) := c(i,i+1) + fs^-_\f(i+1)$,
$fs^+_\f(i) := fs^-_\d(i+1)$\\
$fs^-_\d(i) := c(i,i+1) + fs^+_\f(i+1)$,
$fs^-_\f(i) := fs^+_\d(i+1)$

\item make a swap in $i$ (if available and if $i>\l$):
Then the other player has to return to $\l$ to avoid a violation of (R2).
The feasibility of such a turn depends on the number of edges between $\l$ and $i$.

If $i-\l$ is even:\\
$fs^+_\d(i) := sw_\d(i)+\mbox{fpath}_\f(i)$,
$fs^+_\f(i) := sw_\f(i)+\mbox{fpath}_\d(i)$\\
$fs^-_\d(i) := \top$, $fs^-_\f(i) := \top$

If $i-\l$ is odd:\\
$fs^+_\d(i) := \top$, $fs^+_\f(i) := \top$\\
$fs^-_\d(i) := sw_\d(i)+\mbox{fpath}_\f(i)$,
$fs^-_\f(i) := sw_\f(i)+\mbox{fpath}_\d(i)$
\end{enumerate}
As an initialization we start at a vertex with a swap option having the largest index
$k' \leq k$ and insert the values of the above Case 2.\ for $i=k'$.

\medskip
Considering the backward swap we let $\mbox{bpath}_P(i)$ denote the cost of the path
from $i$ towards $\l$.
Now we have to perform the following computation for all values of $j=1,\ldots, \l-1$.
Again we have to take the minimum between the following two cases (if the second case exists)
while we move up from $i=j'+1$ (defined below) to $i=\l$.
\begin{enumerate}
\item move on along the cycle to vertex $i-1$:

$bs^+_\d(i,j) := c(i-1,i) + bs^-_\f(i-1,j)$,
$bs^+_\f(i,j) := bs^-_\d(i-1,j)$\\
$bs^-_\d(i,j) := c(i-1,i) + bs^+_\f(i-1,j)$,
$bs^-_\f(i,j) := bs^+_\d(i-1,j)$

\item make a swap in $i$ (if available and if $i < \l$):
Then the other player has to return to $\l$ to avoid a violation of (R2).
The feasibility of such a turn depends on the number of edges between $i$ and $\l$.

If $\l-i$ is even:\\
$bs^+_\d(i,j) := sw_\d(i)+\mbox{bpath}_\f(i)$,
$bs^+_\f(i,j) := sw_\f(i)+\mbox{bpath}_\d(i)$\\
$bs^-_\d(i,j) := \top$, $bs^-_\f(i,j) := \top$

If $\l-i$ is odd:\\
$bs^+_\d(i,j) := \top$, $bs^+_\f(i,j) := \top$\\
$bs^-_\d(i,j) := sw_\d(i)+\mbox{bpath}_\f(i)$,
$bs^-_\f(i,j) := sw_\f(i)+\mbox{bpath}_\d(i)$
\end{enumerate}
As an initialization we start at a vertex with a swap having the smallest index $j' > j$
and insert the values of the above Case 2.\ for $i=j'$.

\medskip
Next we determine the path from $0$ via $k, k-1,\ldots$ to $\l$,
when no turn arounds are possible any more (Case (ii)).
We consider the decision at vertex $i$ and compute the entries of the
dynamic programming array from $i=\l+1$ on until $i=k$.
In each vertex, the decision maker has to take the minimum between the following two cases
(if the second case exists):
\begin{enumerate}
\item move on along the cycle to vertex $i-1$:

$ad^+_\d(i) := c(i-1,i) + ad^-_\f(i-1)$,
$ad^+_\f(i) := ad^-_\d(i-1)$\\
$ad^-_\d(i) := c(i-1,i) + ad^+_\f(i-1)$,
$ad^-_\f(i) := ad^+_\d(i-1)$

\item make a swap in $i$ (if available):
Then the other player has to continue
moving along the edge $(i-1,i)$ towards $\l$ to avoid a violation of (R2).

$ad^+_\d(i) := sw_\d(i)+ad^+_\d(i-1)$,
$ad^+_\f(i) := sw_\f(i)+c(i-1,i)+ad^+_\f(i-1)$\\
$ad^-_\d(i) := sw_\d(i)+ad^-_\d(i-1)$,
$ad^-_\f(i) := sw_\f(i)+c(i-1,i)+ad^-_\f(i-1)$
\end{enumerate}
Finally, we extend the definition to $i=0$ and compute $ad^\pm_P(0)$
according to case 1.\ with $k$ replacing $i-1$.

Note that all entries of $ad^\pm_P$ are extended to include a parameter $j$
in analogy to $bs^\pm_P$
which has no influence on the definition of the above recursion,
although the array entries for different values of $j$ may differ
due to the different initialization.

As an initialization we have to consider the decision in $\l$ for Case (ii).
The decider can either move on directly to the next cycle resp.\ edge in the connection strip
or make a swap, either by a swap option available in $\l$ from the previous contraction
of $G\setminus G'$ or by entering the backward swap.
Because of (R2) at most one of the two swap variants can be part of the solution.
Thus we have:
\begin{eqnarray}\label{eq:adinit}
ad^+_\d(\l,j) &:=& \min\{\ed2v_\d, sw_\d+\edv_\f,\, bs^+_\d(\l,j)+\edv_\f\},\\
ad^-_\d(\l,j) &:=& \min\{\ed2v_\d, sw_\d+\edv_\f,\, bs^+_\d(\l,j)+\edv_\f\}\nonumber.
\end{eqnarray}
The values for the follower are immediately implied by the selection of the minimum.

\medskip
Now we can turn to the main arrays and determine their values for all values of $i$ by moving backwards
from $i=\l-1$ down to $i=1$.
Let $D\in \{A, B\}$ again be the decision maker in $0$.
We let $\mbox{rpath}_P(i)$ denote the cost of the return path
from $i$ towards $0$.
In any vertex $i$ the decision process has at most three outcomes,
from which the cheapest is chosen.
(Note that cases 1.\ and 2.\ are completely analogous to the earlier problem
of contracting a cycle.)
\begin{enumerate}
\item move on along the cycle to vertex $i+1$:

$rd^+_\d(i):=c(i,i+1)+rd^-_\f(i+1)$, $rd^+_\f(i)=rd^-_\d(i+1)$\\
$rd^-_\d(k):=c(i,i+1)+rd^+_\f(i+1)$, $rd^-_\f(i)=rd^+_\d(i+1)$\\
$td^+_\d(i):=c(i,i+1)+td^-_\f(i+1)$, $td^+_\f(i)=td^-_\d(i+1)$\\
$td^-_\d(k):=c(i,i+1)+td^+_\f(i+1)$, $td^-_\f(i)=td^+_\d(i+1)$

\item make a swap (if available):
Then the other player has (at most) two possibilities and chooses the one
with the lower cost between 2a.\ and 2b.\ (or the only feasible choice),
which automatically implies the cost for the decider in $i$.
\item[2a.] move on to vertex $i+1$:

$rd^+_\f(i)=sw_\f(i)+c(i,i+1)+rd^+_\f(i+1)$, $rd^+_\d(i):=sw_\d(i)+ rd^+_\d(i+1)$ \\
$rd^-_\f(i)=sw_\f(i)+c(i,i+1)+rd^-_\f(i+1)$, $rd^-_\d(i):=sw_\d(i)+ rd^-_\d(i+1)$ \\
$td^+_\f(i):= sw_\f(i) + c(i,i+1)+ rd^+_\f(i+1)$, $td^+_\d(i):= sw_\d(i) + rd^+_\d(i+1)$\\
$td^-_\f(i):= sw_\f(i) + c(i,i+1)+ rd^-_\f(i+1)$, $td^-_\d(i):= sw_\d(i) + rd^-_\d(i+1)$.

\item[2b.] turn around (if possible):
Then the players have to go directly back to $0$ at cost $\mbox{rpath}_P(i)$
and then move to $\l$ at cost $ad^\pm_P(0,i)$ (Case (ii)).
Recall that the second parameter of $ad$, $i$ in our case, indicates a lower bound on the
end vertex of a possible backward swap.
Since the decider in vertex $0$ must be different from $D$, the feasibility of a turn around in $i$ depends on the number of edges between $0$ and $i$.
This yields:\\
If $i$ is even:\\
$td^+_\d(i):= sw_\d(i) + \mbox{rpath}_\f(i)+ad^-_\f(0,i)$,\\
$td^+_\f(i):= sw_\f(i) + \mbox{rpath}_\d(i)+ad^-_\d(0,i)$,\\
$td^-_\d(i):= \top$, $td^-_\f(i):= \top$

If $i$ is odd:\\
$td^+_\d(i):= \top$, $td^+_\f(i):= \top$\\
$td^-_\d(i):= sw_\d(i) + \mbox{rpath}_\f(i)+ad^-_\d(0,i)$,\\
$td^-_\f(i):= sw_\f(i) + \mbox{rpath}_\d(i)+ad^-_\f(0,i)$
\end{enumerate}

For the initialization we have to consider the decision in $\l$ for Case (i).
Note that for Case (ii) a decision was made in some vertex $j$ to turn around.
For this decision the cost of the full path from $j$ back to $0$ and then
to the next cycle and onwards to $t$ had to be taken into account.
These costs were included in the initialization of $ad^\pm_P$ in (\ref{eq:adinit}).
For Case (i) we proceed in a similar way:
The decider can either move on directly to the next cycle resp.\ edge in the connection strip
or make a swap, either by a swap option available in $\l$ from the previous contraction
of $G\setminus G'$ or by entering the forward swap.
Because of (R2) at most one of the two swap variants can be part of the solution.
After a swap the other player has to move on to the next cycle resp.\ edge.
Thus we have: \\
$rd^+_\d(\l):= \min\{\ed2v_\d, sw_\d+\edv_\f,\, fs^+_\d(\l)+\edv_\f\}$,\\
$rd^-_\d(\l):= \min\{\ed2v_\d, sw_\d+\edv_\f,\, fs^+_\d(\l)+\edv_\f\}$\\
The values for the follower are immediately implied by the selection of the minimum.

In the case that a turn around is still possible, i.e.\ no swap is performed in any vertex
between $0$ and $\l$, there is also the special case of reaching a swap by
traversing the full cycle and repeating the path from $0$ to $\l$.
Therefore we denote by $fc^\pm_P(i)$ (full cycle) the minimal cost to move from
$i$, $i=\l,\ldots,k$, to $0$ with possible swaps on the way and then via $1,2, \ldots$
to $\l$ without swaps. The parity $\pm$ is aligned with the decider in $\l$.
The computation of $fc^\pm_P$ is very similar to $fs^\pm_P$, but with the addition
of the fixed path $0,1,\ldots, \l$ instead of the return path.
Thus we refrain from giving the details.
The initialization for $i=k$ has to observe the parity condition to guarantee
that the going round the full cycle actually changes the role of the decider in $\l$.
We get:\\
$td^+_\d(\l):= \min\{\ed2v_\d, sw_\d+\edv_\f, fs^+_\d(\l)+\edv_\f, fc^+_\d+\edv_\f\}$,\\
$td^-_\d(\l):= \min\{\ed2v_\d, sw_\d+\edv_\f, fs^+_\d(\l)+\edv_\f, fc^+_\d+\edv_\f\}$

\medskip
At the end of these computations we can determine the exit costs
for the cycle or edge preceding the current cycle,
i.e.\ the costs of the \opath\ from $v_0$ to $t$.
It is easy to see how these exits costs can be propagated along an edge
leading from vertex $v_0$ of the current cycle towards $s$.
The general case of the \opath\ for the current cycle yields
$\edv_\d := td^+_\d(0)$ and
$\edv_\f := td^+_\f(0)$.
If the path for the current cycle is restricted and thus leaves more
freedom of choice for the preceding cycle,
we have to compute $\ed2v_P$ by repeating all computations for the current cycle
with the restriction that $v_0$ can be visited only once.
This means that Case (ii) and the special case of Case (i), which traverses the full cycle,
are simply removed from consideration and the reduced problem
is computed as before.
We refrain from giving the simple details of this process.

Recall that we also have to consider the mirrored situation where $D$ picks
$(0,k)$ as the first edge in the cycle and run all the computations
once more with exactly opposite configurations.
At the end, we determine each of $\edv_\d$ and $\ed2v_\d$
as the minimum among among the two directions with the obvious consequences
for the follower.

\bigskip
For a {\em directed cactus graph}, i.e.\ a directed graph which can be obtained from an
undirected cactus graph by assigning a direction to each edge,
the number of possibilities for the two players
to explore a cycle is much more restricted.
A cycle may be directed in a cyclic order and thus allow a swap option
(in the contraction of branches) or a full cycle in the main part of the algorithm
(Case (i), special case), both in a unique way.
Or a cycle permits two paths from $v_0$ to $v_\ell$ in the main part of the algorithm,
which can be settled by simply exploring both options for the decider in $v_0$.
In this case, no swap is possible and such a cycle can be eliminated
if it is in $G \setminus G'$.
Finally, a cycle may leave only a unique way for its traversal thus being equivalent to a
sequence of edges or contain directions that leave it blocked.
In all cases, we can find the \opath\ in polynomial time
by a considerably simplified version of the above algorithm.

\subsection{Running time}
\label{sec:time}

Let $n:=|V|$.
The overall execution of the algorithm is based on the tree structure
inherent in every cactus graph by contracting its cycles.
The algorithm first considers all subtrees of this tree
lying outside the unique path from $s$ to $t$ and contracts recursively all leaves
of these subtrees to their parent vertex.
Then the main path is resolved moving from $t$ to $s$.
All together, each cycle is contracted into a constant number (2 or 4) of cost values
(swap option or exit costs)
containing all the information of previously considered parts of the graph.
Thus, the overall running time is given by the sum of running times
for the contraction of all cycles in $G'$ and $G\setminus G'$.

Considering the computation for one cycle of the connection strip $G'$,
we can observe that each entry of the dynamic programming arrays
can be computed in constant time.
Entries of the arrays are computed once by moving along certain parts of the cycle
which yields on overall linear running time (linear in the number of vertices of the cycle).
This also applies to the arrays used in the branch contraction of $G\setminus G'$
performed in Section~\ref{sec:contr}.
There are two notable exceptions from this property,
namely the backward swap $bs^\pm_P(i,j)$ and the alternative path $ad^\pm_P(i,j)$
which have a quadratic number of entries to be computed,
each of them in constant time.
Thus, we have the following statement.
\begin{theorem}\label{th:time}
The \opath\ of \spg\ on undirected cactus graphs can be computed in $O(n^2)$ time.
\end{theorem}

Since backward swaps do not apply for directed graphs we can state immediately.
\begin{coro}\label{th:timedir}
The \opath\ of \spg\ on directed cactus graphs can be computed in $O(n)$ time.
\end{coro}

\subsection{More general classes of graphs and some observations}		
\label{sec:moregeneral}

\begin{figure}[htb]
\begin{center}
{
\scalebox{1} 
{
\begin{pspicture}(0,-3.7767189)(8.449062,3.8167188)
\definecolor{color320}{rgb}{0.12549019607843137,0.0,1.0}
\definecolor{color512}{rgb}{0.00392156862745098,0.00392156862745098,0.00392156862745098}
\psellipse[linewidth=0.01,linecolor=color512,linestyle=dashed,dash=0.16cm 0.16cm,dimen=outer](3.78,-2.5467188)(2.25,1.23)
\psline[linewidth=0.024cm,linecolor=color320](1.83,-2.5767188)(5.03,-3.1767187)
\psline[linewidth=0.024cm,linecolor=red](0.41,0.14328125)(1.81,0.14328125)
\psline[linewidth=0.024cm,linecolor=color320](1.81,0.14328125)(3.41,0.74328125)
\psline[linewidth=0.024cm,linecolor=red](3.41,0.74328125)(5.01,0.74328125)
\psline[linewidth=0.024cm,linecolor=color320](5.01,0.74328125)(5.01,-0.45671874)
\psline[linewidth=0.024cm,linecolor=red](5.01,-0.45671874)(7.81,0.14328125)
\psdots[dotsize=0.12](0.41,3.0432813)
\psdots[dotsize=0.12](1.81,3.0432813)
\psdots[dotsize=0.12](3.41,3.6432812)
\psdots[dotsize=0.12](5.01,3.6432812)
\psdots[dotsize=0.12](5.01,2.4432812)
\psdots[dotsize=0.12](7.81,3.0432813)
\psline[linewidth=0.024cm](0.41,3.0432813)(1.81,3.0432813)
\psline[linewidth=0.024cm](1.81,3.0432813)(3.41,3.6432812)
\psline[linewidth=0.024cm](3.41,3.6432812)(5.01,3.6432812)
\psline[linewidth=0.024cm](5.01,3.6432812)(5.01,2.4432812)
\psline[linewidth=0.024cm](5.01,2.4432812)(7.81,3.0432813)
\psline[linewidth=0.024cm](5.01,3.6432812)(7.81,3.0432813)
\psline[linewidth=0.024cm](1.81,3.0432813)(5.01,2.4432812)
\usefont{T1}{ptm}{m}{n}
\rput(0.24453124,3.3532813){$s$}
\usefont{T1}{ptm}{m}{n}
\rput(8.094531,2.8332813){$t$}
\usefont{T1}{ptm}{m}{n}
\rput(6.5445313,3.6132812){$M$}
\usefont{T1}{ptm}{m}{n}
\rput(6.434531,2.4332812){$2$}
\usefont{T1}{ptm}{m}{n}
\rput(5.1945314,2.9932814){$1$}
\psdots[dotsize=0.12](0.41,0.14328125)
\psdots[dotsize=0.12](1.81,0.14328125)
\psdots[dotsize=0.12](3.41,0.74328125)
\psdots[dotsize=0.12](5.01,0.74328125)
\psdots[dotsize=0.12](5.01,-0.45671874)
\psdots[dotsize=0.12](7.81,0.14328125)
\psline[linewidth=0.024cm](5.01,0.74328125)(7.81,0.14328125)
\psline[linewidth=0.024cm](1.81,0.14328125)(5.01,-0.45671874)
\usefont{T1}{ptm}{m}{n}
\rput(0.24453124,0.45328125){$s$}
\usefont{T1}{ptm}{m}{n}
\rput(8.094531,-0.06671875){$t$}
\usefont{T1}{ptm}{m}{n}
\rput(6.6445312,0.63328123){$M$}
\usefont{T1}{ptm}{m}{n}
\rput(6.434531,-0.46671876){$2$}
\usefont{T1}{ptm}{m}{n}
\rput(5.1945314,0.09328125){$1$}
\usefont{T1}{ptm}{m}{n}
\rput(3.5609374,1.4532813){\opath}
\psdots[dotsize=0.12](0.43,-2.5767188)
\psdots[dotsize=0.12](1.83,-2.5767188)
\psdots[dotsize=0.12](3.43,-1.9767188)
\psdots[dotsize=0.12](5.03,-1.9767188)
\psdots[dotsize=0.12](5.03,-3.1767187)
\psdots[dotsize=0.12](7.83,-2.5767188)
\psline[linewidth=0.024cm](0.43,-2.5767188)(1.83,-2.5767188)
\psline[linewidth=0.024cm](1.83,-2.5767188)(3.43,-1.9767188)
\psline[linewidth=0.024cm](3.43,-1.9767188)(5.03,-1.9767188)
\psline[linewidth=0.024cm](5.03,-1.9767188)(5.03,-3.1767187)
\psline[linewidth=0.024cm](5.03,-3.1767187)(7.83,-2.5767188)
\psline[linewidth=0.024cm](5.03,-1.9767188)(7.83,-2.5767188)
\usefont{T1}{ptm}{m}{n}
\rput(0.26453125,-2.2667189){$s$}
\usefont{T1}{ptm}{m}{n}
\rput(8.1145315,-2.7867188){$t$}
\usefont{T1}{ptm}{m}{n}
\rput(6.6445312,-2.0267189){$M$}
\usefont{T1}{ptm}{m}{n}
\rput(6.454531,-3.1867187){$2$}
\usefont{T1}{ptm}{m}{n}
\rput(5.2145314,-2.6267188){$1$}
\usefont{T1}{ptm}{m}{n}
\rput(1.7845312,-0.14671876){$a$}
\usefont{T1}{ptm}{m}{n}
\rput(4.7945313,-0.70671874){$b$}
\usefont{T1}{ptm}{m}{n}
\rput(1.8245312,-2.8667188){$a$}
\usefont{T1}{ptm}{m}{n}
\rput(4.8145313,-3.3867188){$b$}
\end{pspicture}
}

}
\end{center}
\caption{\label{fig:counterex} Example where the connection from $a$ to $b$
used in the global \opath\ from $s$ to $t$ is disjoint form the local \opath\ from $a$ to $b$.
Unlabeled edges have cost $0$.}
\end{figure}
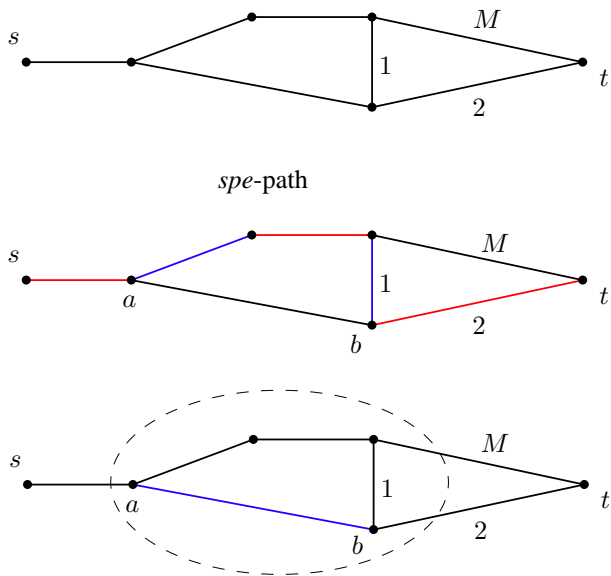

The main reason why \spg\ is solvable in polynomial time on cactus graphs,
which have treewidth $2$,
is the fact that optimal solutions of subgraphs can be used for deriving
an optimal solution of the whole graph.
In this section we give a simple example showing that this does not work anymore
for outerplaner graphs, which still have treewidth $2$ and are a superclass of cactus graphs:
In Figure~\ref{fig:counterex} the path corresponding to the optimal strategy
is illustrated by giving edges chosen by player $1$ the color red
and its opponent blue.
The \opath\ has length $(2,1)$ and goes through vertices $a$ and $b$.
However, when restricting the problem to the subgraph $G\setminus\{s,t\}$
(framed by the dashed line in the third picture of Figure~\ref{fig:counterex})
a \opath\ from $a$ to $b$ with player $2$ starting in $a$ consists of
simply choosing edge $(a,b)$ with length $(0,0)$.
If however player $2$ would choose this edge when playing the game on the whole graph,
player $1$ would choose the edge of length $1$ forcing player $2$ to use
the edge of length $M$ in order to reach $t$.
Thus, the connection from $a$ to $b$ traversed in the global \opath\
is different from the local subgame perfect equilibrium path from $a$ to $b$.

For narrowing the gap between the positive result for cactus graphs (having treewidth $2$)
and the negative result of general (and bipartite) graphs,
it would be interesting to consider graphs of bounded treewidth, as it was done for \geo.
However, a potential \psp ness result for \spg\ on bounded treewidth graphs along the lines of the proof
of Theorem~\ref{the:psp} does not seem to be within reach.
To be in line with that proof,
we would need  a restricted variant of \textsc{Quantified $3$-Sat} satisfying the property of having \textit{respectful} bounded treewidth (cf.~\citet{asol13}) in order to achieve bounded treewidth of the graph used. However, \citet{asol13}  note that bounded treewidth quantified boolean formula become polynomial time solvable when restricted to \textit{respectful} bounded treewidth instances by a result of \citet{chda12}.
Thus, a potential \psp ness result for \spg\ on bounded treewidth graphs will likely require a different approach;
the computational complexity of \spg\ on bounded treewidth graphs remains an interesting open question.


\medskip

\subsection*{Acknowledgements}

We would like to thank Christian Klamler (University of Graz)
for fruitful discussions and valuable comments.
We would also like to thank the anonymous referees for their comments
that helped a lot to improve the presentation of the paper.

Ulrich Pferschy and Joachim Schauer were supported by the Austrian
Science Fund (FWF): {[}P 23829-N13{]}. \
Andreas Darmann was supported by the Austrian Science Fund (FWF):
{[}P 23724-G11{]}.

\bibliographystyle{plainnat}
\bibliography{path-game}

\begin{thebibliography}{15}
\providecommand{\natexlab}[1]{#1}
\providecommand{\url}[1]{\texttt{#1}}
\expandafter\ifx\csname urlstyle\endcsname\relax
  \providecommand{\doi}[1]{doi: #1}\else
  \providecommand{\doi}{doi: \begingroup \urlstyle{rm}\Url}\fi

\bibitem[Atserias and Oliva(2014)]{asol13}
Albert Atserias and Sergi Oliva.
\newblock Bounded-width {QBF} is {PSPACE}-complete.
\newblock \emph{Journal of Computer and System Sciences}, 80\penalty0
  (7):\penalty0 1415 -- 1429, 2014.

\bibitem[Bodlaender(1993)]{bo93}
H.~Bodlaender.
\newblock Complexity of path-forming games.
\newblock \emph{Theoretical Computer Science}, 110\penalty0 (1):\penalty0
  215--245, 1993.

\bibitem[Chen and Dalmau(2012)]{chda12}
H.~Chen and V.~Dalmau.
\newblock Decomposing quantified conjunctive (or disjunctive) formulas.
\newblock In \emph{Proceedings of the 27th Annual ACM/IEEE Symposium on Logic
  in Computer Science, LICS 2012}, pages 205--214, 2012.

\bibitem[Cormen et~al.(2009)Cormen, Leiserson, Rivest, and Stein]{Cormen}
T.H. Cormen, C.E. Leiserson, R.L. Rivest, and C.~Stein.
\newblock \emph{Introduction to Algorithms, 3rd ed.}
\newblock MIT Press, 2009.

\bibitem[Darmann et~al.(2014{\natexlab{a}})Darmann, Klamler, and
  Pferschy]{dkp13}
A.~Darmann, C.~Klamler, and U.~Pferschy.
\newblock Sharing the cost of a path.
\newblock \emph{to appear in: {\sl Studies in Microeconomics}},
  2014{\natexlab{a}}.
\newblock available at:\\ {\tt http://ssrn.com/abstract=2287875}.

\bibitem[Darmann et~al.(2014{\natexlab{b}})Darmann, Nicosia, Pferschy, and
  Schauer]{dnp13}
A.~Darmann, G.~Nicosia, U.~Pferschy, and J.~Schauer.
\newblock The subset sum game.
\newblock \emph{European Journal on Operational Research}, 233:\penalty0
  539--549, 2014{\natexlab{b}}.

\bibitem[Darmann et~al.(2014{\natexlab{c}})Darmann, Pferschy, and
  Schauer]{dps14}
A.~Darmann, U.~Pferschy, and J.~Schauer.
\newblock The shortest path game: Complexity and algorithms.
\newblock In \emph{Proceedings of Theoretical Computer Science (TCS 2014),
  Rome}, volume 8705 of \emph{Lecture Notes in Computer Science}, pages 39--53.
  Springer, 2014{\natexlab{c}}.

\bibitem[Fraenkel and Goldschmidt(1987)]{goldi}
A.S. Fraenkel and E.~Goldschmidt.
\newblock {PSPACE}-hardness of some combinatorial games.
\newblock \emph{Journal of Combinatorial Theory}, 46\penalty0 (1):\penalty0
  21--38, 1987.

\bibitem[Fraenkel and Simonson(1993)]{frsi93}
A.S. Fraenkel and S.~Simonson.
\newblock Geography.
\newblock \emph{Theoretical Computer Science}, 110\penalty0 (1):\penalty0
  197--214, 1993.

\bibitem[Fraenkel et~al.(1993)Fraenkel, Scheinerman, and Ullman]{fsu93}
A.S. Fraenkel, E.R. Scheinerman, and D.~Ullman.
\newblock Undirected edge geography.
\newblock \emph{Theoretical Computer Science}, 112\penalty0 (2):\penalty0
  371--381, 1993.

\bibitem[Lichtenstein and Sipser(1980)]{lisi80}
D.~Lichtenstein and M.~Sipser.
\newblock Go is polynomial-space hard.
\newblock \emph{Journal of the ACM}, 27\penalty0 (2):\penalty0 393--401, 1980.

\bibitem[Nisan et~al.(2007)Nisan, Roughgarden, Tardos, and (eds.)]{nrt07}
N.~Nisan, T.~Roughgarden, E.~Tardos, and V.V.~Vazirani (eds.).
\newblock \emph{Algorithmic Game Theory}.
\newblock Cambridge University Press, 2007.

\bibitem[Osborne(2004)]{osb04}
M.J. Osborne.
\newblock \emph{An Introduction to Game Theory}.
\newblock Oxford University Press, USA, 2004.

\bibitem[Schaefer(1978)]{sch78}
T.J. Schaefer.
\newblock On the complexity of some two-person perfect-information games.
\newblock \emph{Journal of Computer and System Sciences}, 16\penalty0
  (2):\penalty0 185--225, 1978.

\bibitem[Stockmeyer and Meyer(1973)]{stocki}
L.J. Stockmeyer and A.R. Meyer.
\newblock Word problems requiring exponential time.
\newblock In \emph{Proceedings of the 5th Symposium on Theory of Computing,
  STOC '73}, pages 1--9. ACM, 1973.

\end{thebibliography}

\end{document}